\pdfoutput=1
\documentclass[prd,onecolumn,nofootinbib,superscriptaddress]{revtex4-2}
\usepackage{graphicx}
\usepackage{bm}
\usepackage[colorlinks=true]{hyperref}
\usepackage[all]{hypcap}
\usepackage{float}
\usepackage{amsmath}
\usepackage{amssymb}
\usepackage{mathrsfs}
\usepackage{pifont}
\usepackage{gensymb}
\usepackage{multirow}
\usepackage[dvipsnames]{xcolor}
\usepackage[T1]{fontenc}
\usepackage[utf8]{inputenc}
\usepackage{orcidlink}
\usepackage[normalem]{ulem}
\usepackage{tikz,pgfplots}
\usetikzlibrary{shapes.geometric,arrows.meta,positioning,matrix,decorations.markings,snakes,external}
\tikzset{>=Stealth}

\ifdefined\myext
\fi

\def\relsstandalone{}

\newcommand\be{\begin{equation}}
\newcommand\ba{\begin{eqnarray}}
\newcommand\ee{\end{equation}}
\newcommand\ea{\end{eqnarray}}

\newcommand\bw{\begin{widetext}}
\newcommand\ew{\end{widetext}}

\newcommand{\avg}[1]{\left \langle #1 \right \rangle}
\newcommand{\cd}{\nabla}

\newcommand{\bs}[1]{\boldsymbol{#1}}
\newcommand{\ud}{\mathrm{d}}
\newcommand{\lie}{\pounds}
\newcommand{\mc}[1]{\mathcal{#1}}
\newcommand{\ord}[2]{\underset{^{(#1)}}{#2}{}}
\newcommand{\interchange}[2]{#1 \longleftrightarrow #2}
\newcommand{\scri}{\mathscr{I}}

\makeatletter
\newcommand\footnoteref[1]{\protected@xdef\@thefnmark{\ref{#1}}\@footnotemark}
\makeatother
\definecolor{KellyGreen}{RGB}{76,187,23}

\newcommand{\figLengths}{%
\begin{figure}
  \ifdefined\myext
    \tikzsetnextfilename{Wavelength}
    \resizebox{0.95\textwidth}{!}{\input{Wavelength.tikz}}
  \else
    \ifx\relsstandalone\undefined
      \resizebox{0.95\textwidth}{!}{\input{Wavelength.tikz}}
    \else
      \includegraphics[width=0.95\textwidth]{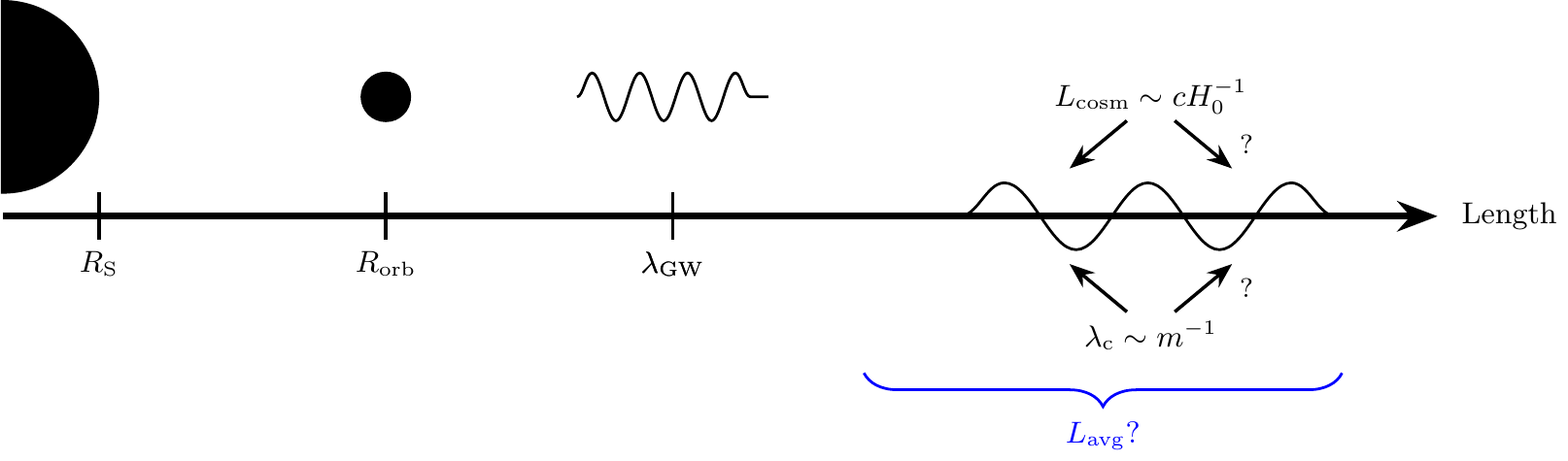}
    \fi
  \fi
  \caption{Length scale comparison for a typical binary system in bigravity. 
  In GR, there is a clear hierarchy of length scales $R_S \ll R_{\text{orb}} \ll \lambda_{\text{GW}} \ll L_\text{ST}$. 
  There is an unambiguous choice for $L_\text{avg}$ to separate the last two, $\lambda_\text{GW} \ll L_\text{avg} \ll L_\text{ST}$.
  Bigravity introduces additional length scales, without an obvious hierarchy, so the 
  choice of averaging length scale is ambiguous.
  \label{fig:Wavelength}}
\end{figure}
}

\newcommand{\figRadiation}{%
\begin{figure}
  \ifdefined\myext
    \tikzsetnextfilename{Radiation}
    \resizebox{0.5\textwidth}{!}{\input{Radiation.tikz}}
  \else
    \ifx\relsstandalone\undefined
      \resizebox{0.5\textwidth}{!}{\input{Radiation.tikz}}
    \else
      \includegraphics[width=0.5\textwidth]{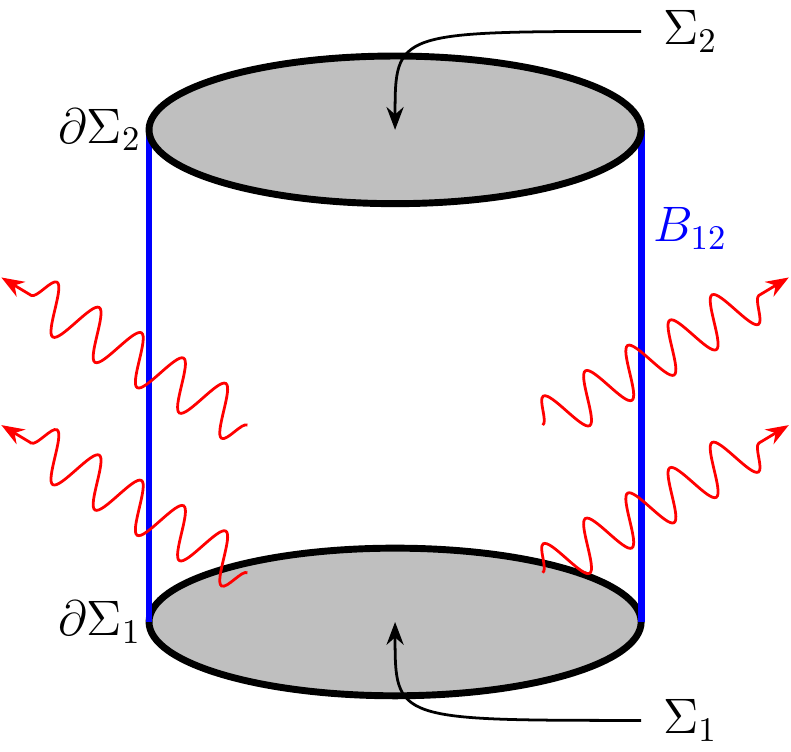}
    \fi
  \fi
  \caption{Geometry of various approaches to finding conserved quantities.  Integration about $\Sigma_{1,2}$ will present the amount of a conserved quantity.  Integrating over the boundary $B_{12}$ can yield the flux of this conserved quantity.  Lastly, integration over the boundary terms $\partial \Sigma_{12}$, allows for the calculation of asymptotic charges.}
  \label{fig:Radiation}
\end{figure}
}

\newcommand{\guelph}{Department of Physics, University of Guelph, 50 Stone Road E., Guelph, ON N1G 2W1, Canada}
\newcommand{\PI}{Perimeter Institute for Theoretical Physics, Waterloo, ON N2L 2Y5, Canada}

\begin{document}
\title{Gravitational-wave energy and other fluxes in ghost-free bigravity}

\author{Alexander M. Grant \orcidlink{0000-0001-5867-4372}}
\email{alex.grant@virginia.edu}
\affiliation{Department of Physics, University of Virginia, Charlottesville, Virginia 22904, USA}

\author{Alexander Saffer \orcidlink{0000-0001-7832-9066}}
\email{alexander.saffer@mail.wvu.edu}
\affiliation{Department of Physics, University of Virginia, Charlottesville, Virginia 22904, USA}
\affiliation{Department of Physics and Astronomy, West Virginia University, P.O. Box 6315, Morgantown, WV 26506, USA}
\affiliation{Center for Gravitational Waves and Cosmology, West Virginia University, Chestnut Ridge Research Building, Morgantown, WV 26505,
USA}

\author{Leo C. Stein	\orcidlink{0000-0001-7559-9597}}
\email{lcstein@olemiss.edu}
\affiliation{Department of Physics and Astronomy,
		University of Mississippi, University, Mississippi 38677, USA}

\author{Shammi Tahura \orcidlink{0000-0001-5678-5028}}
\email{stahura@uoguelph.ca}
\affiliation{Department of Physics, University of Virginia, Charlottesville, Virginia 22904, USA}
\affiliation{\guelph}
\affiliation{\PI}

\date{\today}
\begin{abstract}
One of the key ingredients for making binary waveform predictions in a beyond-GR theory of gravity is understanding the energy and angular momentum carried by gravitational waves and any other radiated fields.
Identifying the appropriate energy functional is unclear in Hassan-Rosen bigravity, a ghost-free theory with one massive and one massless graviton.
The difficulty arises from the new degrees of freedom and length scales which are not present in GR, rendering an Isaacson-style averaging calculation ambiguous.
In this article we compute the energy carried by gravitational waves in bigravity starting from the action, using the canonical current formalism.
The canonical current agrees with other common energy calculations in GR, and is unambiguous (modulo boundary terms), making it a convenient choice for quantifying the energy of gravitational waves in bigravity or any diffeomorphism-invariant theories of gravity.
This calculation opens the door for future waveform modeling in bigravity to correctly include backreaction due to emission of gravitational waves.
\end{abstract}

\maketitle

{
    \hypersetup{linkcolor=blue}
    \tableofcontents
}

\section{Introduction}

The study of massive spin-2 particles can be traced back to the work of Fierz and Pauli~\cite{Fierz:1939ix}.  Their work was the first to construct a linear solution to the relativistic wave equation with the addition of a massive spin-2 field. 
Later, Salam et al. added a massive spin-2 field with a massless graviton demonstrating the use of a dynamical reference metric and expanding on the theory of massive gravity~\cite{Isham:1971gm}.
However, attempts to further the theory to a non-linear regime failed due to the presence of a ghost-like scalar propagating mode~\cite{Boulware:1972yco}.  It would not be for several more decades before de Rham and Gabadadze showed that by using an effective field theory (EFT) approach, specific coefficients can be chosen such that the decoupling limit is ghost-free~\cite{deRham:2010ik}.
This work led to the formulation of de Rham-Gabadadze-Tolley theory, where the action was properly generalized to remove additional scalar ghosts~\cite{deRham:2010kj,deRham:2014zqa}.  
This theory's breakthrough was to introduce the proper interaction between the two metrics in the Lagrangian which allowed for a closed form, non-linear action for a massive gravity theory without ghosts.
Following this work, Hassan and Rosen promoted the auxiliary metric to a dynamical field~\cite{Hassan:2011vm,Hassan:2011hr,Hassan:2011tf,Hassan:2011zd,Hassan:2011ea}.  This led to a ghost-free theory with seven degrees of freedom.  Massive bigravity (or bimetric gravity depending on your source)~\cite{Schmidt-May:2015vnx}, is the resulting theory and the one we are interested in with this work.

Previous studies of this theory in astrophysical settings have resulted in black hole~\cite{Comelli:2011wq,Volkov:2012wp,Babichev:2014oua,Ayon-Beato:2015qtt, Brito:2013xaa}, neutron star~\cite{Sullivan:2017kwo}, exact plane wave~\cite{Mohseni:2012ug, Ayon-Beato:2018hxz}, and cosmological solutions~\cite{Volkov:2011an,vonStrauss:2011mq,Comelli:2011zm,Enander:2013kza}.  However, these works showed there is no analogue to Birkhoff's theorem for bigravity~\cite{Schmidt-May:2015vnx, Kocic:2017hve}.  Particularly for the case of black holes, this leads to the property that these objects might have ``hair''~\cite{Volkov:2012wp,Brito:2013xaa}, the presence of which would leave an imprint on the quasi-normal mode frequencies within the ring-down regime of a binary waveform~\cite{Dreyer:2003bv}.  The past several years of gravitational wave (GW) detections have allowed for bounds to be placed on the ability of black holes to have hair~\cite{Isi:2019aib,2021arXiv211206861T}.
There are, moreover, cosmological constraints on bigravity, see for example~\cite{vonStrauss:2011mq, Enander:2013kza, Hogas:2021lns, Hogas:2021saw}.

 Although bigravity theories are usually considered to explain the late-time cosmic expansion, there have been studies regarding bigravity in the context of dark matter as well~\cite{Aoki:2016zgp}. There are instabilities in linear cosmological perturbations in bigravity theories, which can be remedied by adding a dark sector or considering special initial conditions or backgrounds~\cite{Aoki:2016zgp,Koennig:2014ods,Lagos:2014lca,Comelli:2012db}. Furthermore, there are viable extensions of bigravity to chameleon bigravity to make massive gravity compatible with solar system tests without using the Vainshtein mechanism~\cite{DeFelice:2017oym}. A constrained Hamiltonian theory of ghost-free bigravity that contains four degrees of freedom instead of seven has been obtained as well~\cite{DeFelice:2020ecp}.
 For simplicity we only consider the Hassan-Rosen massive bigravity theory, though it should be straightforward to apply the techniques of this paper to other bigravity theories.

The presence of additional degrees of freedom for our system could lead to a gravitational radiation flux for a binary system that deviates from that expected in general relativity (GR)~\cite{Will:1993ns,Will:2014kxa}.  In this paper, we will concern ourselves with the study of energy flux in bigravity.  This is often found by first calculating an effective gravitational-wave stress energy tensor and integrating this quantity over a constant-time cross-section at null infinity.  A variety of methods for finding the energy-momentum tensor have been found in GR, but some of the most common methods used in GR include the ``Isaacson'' method~\cite{Isaacson:1968hbi,Isaacson:1968zza}, method of a perturbed action~\cite{Stein:2010pn}, finding a pseudo-tensor through the use of tensor densities to construct a conserved quantity based on symmetry~\cite{Landau:1975pou}, as well as using diffeomorphism invariance of a theory to build a conserved quantity~\cite{Noether_1971}.

The methods above have several drawbacks when attempting to apply them to bigravity.  For one, the methods above place little physical meaning on the quantities when applied to a theory where two dynamical metrics are present.  For example, given the Isaacson Method, we find ourselves with two separate field equations, but how could a ``proper'' source of the gravitational curvature be separated out to call it the typical ``stress-energy tensor.''  This is more straightforward in GR, but convoluted when it is unknown how two metrics will act in general.  A second concern which arises is the issue of averaging, which Isaacson performed in order to eliminate the higher frequency variations in a gravitational wave, as well as to obtain a conserved current which was gauge-invariant. With the current understanding of bigravity, the characteristic length over which an averaging procedure needs to take place is not clear (see Fig.~\ref{fig:Wavelength}).  With the characteristic curvature of two separate metrics, as well as the wavelength of the graviton and any propagating GW wavelength, finding an averaging procedure to damp out the highly oscillatory terms is not trivial since how these lengths compare to one another is not known a priori.

A second method of investigating the energy flux can be through the use of a canonical current~\cite{Lee1990, Burnett1990}. These currents are locally conserved and associated with specific Killing symmetries, and have been used for a variety of applications in the literature~\cite{Friedman1978, Hollands2012, Prabhu2015a, Prabhu2016, Sorce2017}.  Should we be interested in the flux of energy of our system, we should focus our efforts with a temporal Killing field.  However, this formulation will also allow for linear and momentum flux based on translational or rotational Killing fields as well~\cite{Bonga2019}. The method of finding these conserved currents involves the variation of the Lagrangian with respect to the different dynamical fields considered in the problem.  In this work, we derive for the first time a conserved current in bigravity theory. One aspect of note is the fact that although the theory was motivated with the assumption that there exists a massive graviton, the resulting quantity $m_{\rm FP}$ does not appear in the final result.  An analogous situation appears in the case of a massive scalar field, which we discuss in Appendix~\ref{app:scalar}.

The paper is outlined as follows: in Sec.~\ref{sec:BigravReview} we present an overview of ghost-free massive bigravity in both a general and linearized form.  Section~\ref{sec:currents} discusses our attempts at constructing conserved quantities.  Here, we make use of the Isaacson approach~\ref{sec:isaacson}, method of canonical currents~\ref{sec:canonical}, as well as an overview of asymptotic charges~\cite{Iyer1994,Wald:1999wa} in Sec.~\ref{sec:asyCharge}.
We use the  the conventions for the metric,  Riemann tensor, and differential forms from Wald~\cite{Wald1984}.
We work in geometric units, where $G = c = 1$, but $\hbar \neq 1$.
For simplicity of notation for our various metrics, quantities pertaining to the $g_{ab}$ metric will be written as standard notation while those relating to $f_{ab}$ will be denoted with an overhead tilde ($\,\tilde{}\,$).
While using an index-free (or partially index-free) notation for tensors, we denote these tensors in bold (this is standard for differential forms; we merely extend this convention to all tensors).
In particular, we use bold for tensors that appear as the arguments of functions and functionals, except for the determinant.

\section{Review of Bigravity}
\label{sec:BigravReview}
\subsection{General formulation}

The theory of ghost-free massive bigravity is discussed in~\cite{Hassan:2011zd}, and we present their results here.
We start with the Lagrangian density for the theory, which (using the formalism laid out in e.g.~\cite{Lee1990}) we write as a four-form $\bs L$\footnote{Note that this expression is missing a conventional factor of $\frac{1}{16\pi}$ that typically appears; this is only for brevity, and can easily be added to any of the expressions that appear in this paper.}
\begin{equation} \label{eq:BigravLagrangian}
    \bs L = R \bs \epsilon + \alpha^2 \tilde{R} \tilde{\bs \epsilon} - 2 m^2 V \bs \epsilon,
\end{equation}
where the theory contains two dynamical metrics $g_{ab}$ and $f_{ab}$.
The four-form $\bs \epsilon$ denotes the volume form defined with respect to the metric $g_{ab}$, and $R$ denotes the Ricci scalar constructed from $g_{ab}$---as remarked in the Introduction, quantities with tildes are associated with $f_{ab}$, and so $\tilde{\bs \epsilon}$ is the volume form defined with respect to $f_{ab}$ and $\tilde R$ the Ricci scalar constructed from $f_{ab}$.

The potential $V$ is constructed in such a way as to eliminate the Boulware-Deser ghost~\cite{deRham:2010kj,Bernard:2015uic}:
\begin{equation} \label{eq:BigravPotential}
    V = \sum_{n = 0}^4 \beta_n e_n (\bs S),
\end{equation}
where $\beta_n$ are dimensionless interaction parameters\footnote{Note that, by rescaling these parameters, one can eliminate $m$; similarly, in the case where $f_{ab}$ does not couple to matter, one can eliminate $\alpha$.}, and $e_n (\bs S)$ are the elementary symmetric polynomials constructed from the eigenvalues of the tensor $S^a{}_b$:
\begin{equation}
    \det(\mathbb{I} + t\bs{S}) \equiv \sum_{n = 0}^\infty t^n e_n (\bs S).
\end{equation}
Equivalently, they are given by the tensorial expression~\cite{Kocic:2018ddp}
\begin{equation}
    e_n (\bs S) \equiv S^{[a_1}{}_{a_1} \cdots S^{a_n]}{}_{a_n}.
\end{equation}
It follows from this expression that $e_4 (\bs S) = \det S$ and $e_n (\bs S) = 0$ for $n > 4$.
The tensor $S^a{}_b$ is defined to be the ``square root'' of the tensor $g^{ac} f_{cb}$:
\begin{equation} \label{eq:SDef}
    S^{a}{}_{c} S^{c}{}_{b} = g^{ac} f_{cb}.
\end{equation}
We assume $g, f$ are both invertible, and that the principal square root $S$ exists, which requires that the null cones intersect, admitting a common timelike direction and a hypersurface that is spacelike with respect to both metrics~\cite{Hassan:2017ugh}; then $S^{-1}$ must also exist.
Evaluating the elementary symmetric polynomials for $(S^{-1})^a{}_b$ yields
\begin{equation} \label{eq:invS}
    e_n (\bs S^{-1}) = \frac{e_{4 - n} (\bs S)}{e_4 (\bs S)},
\end{equation}
and since $\det S = \sqrt{-\det f}/\sqrt{-\det g}$, it follows that
\begin{equation}
    e_n (\bs S^{-1}) \tilde{\bs \epsilon} = e_{4 - n} (\bs S) \bs \epsilon.
\end{equation}
This implies that the Lagrangian four-form in Eq.~\eqref{eq:BigravLagrangian} is invariant under the discrete transformation
\begin{equation}
    \interchange{(\alpha^{-1} g_{ab},\, \alpha^{4 - n} \beta_n)}{(\alpha f_{ab},\, \alpha^n \beta_{4 - n})}.
\end{equation}

Finally, note that $m$ has units of inverse length in units where $G = c = 1$, but $\hbar \neq 1$ (as is typically done in classical general relativity). As this form is conventional, we will not introduce a factor of $\hbar$ into the equations which appear in this paper.

Varying the Lagrangian and neglecting total derivative terms (see Sec.~\ref{sec:canonical} for more details on these terms, which are quite important to the construction of conserved currents) we obtain a set of field equations which we express as
\begin{subequations} \label{eq:FieldEquations}
    \begin{align}
        G^{ab} + m^2 \mc V^{ab} &= 0, \\
        \tilde G^{ab} + \frac{m^2}{\alpha^2} \tilde{\mc V}^{ab} &= 0,
    \end{align}
\end{subequations}
where $G^{ab}$ is the standard Einstein tensor constructed from $g_{ab}$, whose indices are raised with $g^{ab}$; similarly, $\tilde G^{ab}$ is constructed from $f_{ab}$ and has its indices raised using $f^{ab}$.
The ``potential'' terms $\mc V^{ab}$ and $\tilde{\mc V}^{ab}$ are defined by
\begin{equation} \label{eq:Vtensors}
    \delta (V \bs \epsilon) = 2 \left(\mc V^{ab} \delta g_{ab} \bs \epsilon + \tilde{\mc V}^{ab} \delta f_{ab} \tilde{\bs \epsilon}\right).
\end{equation}
The details of deriving Eq.~\eqref{eq:FieldEquations} are given in Sec.~\ref{sec:canonical} below.
The tensors $\mc V^{ab}$ and $\tilde{\mc V}^{ab}$ can be written in terms of two tensor fields, $\mc U^a{}_b (\bs S)$ and $\tilde{\mc U}^a{}_b (\bs S)$, which only depend on either metric through $\bs S$:
\begin{equation} \label{eqn:V_from_U}
    \mc V^{ab} = \mc U^a{}_c (\bs S) g^{bc}, \qquad \tilde{\mc V}^{ab} = \tilde{\mc U}^a{}_c (\bs S) f^{bc}
\end{equation}
These tensors, in turn, can be expressed in terms of the elementary symmetric polynomials as
\begin{subequations} \label{eq:UtensorSymPoly}
    \begin{align}
        \mc U^a{}_b (\bs S) &= \sum_{n = 0}^3 (-1)^n \beta_n (Y_n)^a{}_b (\bs S), \\
        \tilde{\mc U}^a{}_b (\bs S) &= \sum_{n = 0}^3 (-1)^n \beta_{4 - n} (Y_n)^a{}_b (\bs S^{-1}),
    \end{align}
\end{subequations}
where the tensors $(Y_n)^a{}_b (\bs S)$ are defined by
\begin{equation} \label{eq:YforSymPoly}
    (Y_n)^a{}_b (\bs S) \equiv \sum_{k = 0}^n (-1)^k e_k (\bs S) (S^{n - k})^a{}_b.
\end{equation}
An alternate definition is given by the variation of the elementary symmetric polynomials \{note the factor of $(-1)^n$ difference from Eq.~(1.44) of~\cite{Kocic:2018ddp}, which arises from a difference in the definition of $(Y_n)^a{}_b (\bs S)$\}:
\begin{equation} \label{eqn:delta_e}
    \delta [e_{n + 1} (\bs S)] = (-1)^n (Y_n)^a{}_b \delta S^b{}_a. 
\end{equation}
Another interesting property of these tensors is the fact that $(Y_n)^a{}_b = 0$ in $n$ dimensions is equivalent to the Cayley-Hamilton theorem~\cite{Edgar2001}.
The exact derivation of this form of $\mc V^{ab}$ and $\tilde{\mc V}^{ab}$ is presented in detail in, for example,~\cite{Bernard:2015mkk}.

By the diffeomorphism invariance of this theory, one can derive a divergence identity given by~\cite{Damour:2002ws}
\begin{equation} \label{eqn:div_identity}
    g_{ab} \nabla_{c} \mc V^{bc} \bs \epsilon + f_{ab} \tilde \nabla_{c} \tilde{\mc V}^{bc} \tilde{\bs \epsilon} = 0.
\end{equation}
In addition to this, Eqs.~\eqref{eq:Vtensors} and~\eqref{eq:SDef} can be used to derive the following identity~\cite{Volkov:2012wp, Hassan:2014vja}:
\begin{equation}
    g_{ab} \mc V^{bc} \bs \epsilon + f_{ab} \tilde{\mc V}^{bc} \tilde{\bs \epsilon} - V \delta_a{}^c \bs \epsilon = 0.
\end{equation}

\subsection{Linearized theory} \label{subsec:BiGravLinearizedMetrics}

We now consider a linearization of our metrics about a general background solution; following~\cite{Lee1990}, we use the same notation as we used for variations above: consider a one-parameter family of metrics $g_{ab} (\lambda)$ and $f_{ab} (\lambda)$, and define the variation operation $\delta$ for any tensor $\bs Q(\lambda)$ by
\begin{equation}
    \delta \bs Q \equiv \left.\frac{\partial \bs Q}{\partial \lambda}\right|_{\lambda = 0}.
\end{equation}
Analogously, in the case where one has an $n$-parameter family (with parameters $\lambda_1, \ldots, \lambda_n$), one can define a variation with respect to any subset of these parameters:
\begin{equation} \label{eqn:multi_var}
    \delta_{i_1} \cdots \delta_{i_j} \bs Q \equiv \left.\frac{\partial^j \bs Q}{\partial \lambda_{i_1} \cdots \partial \lambda_{i_j}}\right|_{\lambda_1 = \cdots = \lambda_n = 0}.
\end{equation}
For simplicity, we denote the background value by $\bs Q$.
Moreover, when computing higher variations, we write
\begin{equation} \label{eqn:repeated_var}
    \delta^n \bs Q \equiv \frac{1}{n!} \left.\frac{\partial^n \bs Q}{\partial \lambda^n}\right|_{\lambda = 0};
\end{equation}
the generalization to the multi-parameter family case is analogous to Eq.~\eqref{eqn:multi_var}.

While the above formalism will be very useful in Sec.~\ref{sec:canonical} below, we specialize in this section to the case of only a single variation, and a one-parameter family.
If one has (in general) nonlinear functional $\bs Q[\bs \Phi]$, where $\bs \Phi$ is some tensor field, then one can construct a linear functional by considering a one-parameter family $\bs \Phi(\lambda)$ and taking the variation of $\bs Q$:
\begin{equation} \label{eqn:ord_1}
    \delta \bs Q[\bs \Phi] \equiv \ord{1}{\bs Q} \{\delta \bs \Phi\}.
\end{equation}
The underset number here indicates the order of the expansion, which in this case is linear.
Here, we denote \emph{explicitly (multi)linear} functionals by curly braces, and reserve square brackets for arbitrary functionals.
Similarly, for the case of a nonlinear \emph{function} $\bs Q(\bs \Phi)$, we write instead
\begin{equation}
    \delta \bs Q(\bs \Phi) \equiv \ord{1}{\bs Q} \cdot \delta \bs \Phi,
\end{equation}
where the ``$\cdot$'' indicates contraction with the indices of $\delta \bs \Phi$.
Except in cases where it is particularly relevant, we suppress the (potentially nonlinear) dependence on $\bs \Phi$, keeping only the linear dependence as explicit: the above equation is an example of this, as (in general) $\ord{1}{\bs Q}$ will depend on the background value $\bs \Phi$.

We now apply this notation to the problem at hand.
Taking a variation of Eq.~\eqref{eq:FieldEquations}, we find that
\begin{subequations} \label{eq:FieldEquations_Linearized}
    \begin{align}
        \ord{1} G^{ab} \{\delta \bs g\} + m^2 \delta \mc V^{ab} &= 0, \label{eq:FEeq1} \\
        \ord{1}{\tilde G}^{ab} \{\delta \bs f\} + \frac{m^2}{\alpha^2} \delta \tilde{\mc V}^{ab} &= 0. \label{eq:FEeq2}
    \end{align}
\end{subequations}
Using Eq.~\eqref{eqn:V_from_U}, we find that
\begin{subequations} \label{eqn:var_V}
    \begin{align}
        \delta \mc V^{ab} &= -\mc V^{ac} g^{bd} \delta g_{cd} + g^{bc} \ord{1}{\mc U}^a{}_{cd}{}^e \delta S^d{}_e, \\
        \delta \tilde{\mc V}^{ab} &= -\tilde{\mc V}^{ac} f^{bd} \delta f_{cd} + g^{bc} \ord{1}{\tilde{\mc U}}^a{}_{cd}{}^e \delta S^d{}_e.
    \end{align}
\end{subequations}
We do not expand these results further, as the formulae for the $\ord{1}{\mc U}^a{}_{cd}{}^e$, $\ord{1}{\tilde{\mc U}}^a{}_{cd}{}^e$, and $\delta S^a{}_b$ are all quite complicated.
However, as this expression will be useful below, we write out the variation of $(Y_n)^a{}_b$: using Eq.~\eqref{eq:YforSymPoly} and~\eqref{eqn:delta_e}, we find that
\begin{equation} \label{eqn:var_Y}
    \delta [(Y_n)^a{}_b (\bs S)] = \sum_{k = 0}^n (-1)^k \left[(-1)^{k - 1} (Y_{k - 1})^d{}_c (\bs S) (S^{n - k})^a{}_b + e_k (\bs S) \sum_{j = 0}^{n - k - 1} (S^j)^a{}_c (S^{n - k - j - 1})^d{}_b\right] \delta S^c{}_d.
\end{equation}

As the general case is rather complicated, we consider only the case where the background solutions are proportional to one another~\cite{Hassan:2012wr},
\begin{equation}
    f_{ab} = c^2 g_{ab},
\end{equation}
where $c$ is a constant (not to be confused with the speed of light, which is 1 in our units).
Proportionality of $f_{ab}$ and $g_{ab}$ implies that
\begin{equation}
    f^{ab} = c^{-2} g^{ab}, \qquad S^a{}_b = c \delta^a{}_b, \qquad \tilde{\bs \epsilon} = c^4 \bs \epsilon,
\end{equation}
whereas the constancy of $c$ implies that
\begin{equation}
    \tilde \nabla_a = \nabla_a, \qquad \tilde G^{ab} = c^{-4} G^{ab}, \qquad \ord{1}{\tilde G}^{ab} \{\bs h\} = c^{-6} \ord{1} G^{ab} \{\bs h\},
\end{equation}
for any rank two tensor $h_{ab}$.
Moreover, we have that
\begin{equation} \label{eqn:e_prop}
    e_n (\bs S^{\pm 1}) = c^{\pm n} \binom{4}{n},
\end{equation}
so that
\begin{equation} \label{eqn:Y_prop}
    (Y_n)^a{}_b (\bs S^{\pm 1}) = c^{\pm n} \sum_{k = 0}^n (-1)^k \binom{4}{k} \delta^a{}_b = (-1)^n c^{\pm n} \binom{3}{n} \delta^a{}_b.
\end{equation}
These equations, together with Eq.~\eqref{eqn:var_Y}, then imply that
\begin{equation} \label{eqn:var_Y_prop}
    \begin{split}
        \delta [(Y_n)^a{}_b (\bs S^{\pm 1})] &= c^{\pm (n - 1)} \sum_{k = 0}^n (-1)^k \left[\binom{3}{k - 1} \delta^d{}_c \delta^a{}_b + (n - k) \binom{4}{k} \delta^a{}_c \delta^d{}_b\right] \delta (S^{\pm 1})^c{}_d \\
        &= (-1)^n c^{\pm (n - 1)} \binom{2}{n - 1} (\delta^d{}_c \delta^a{}_b - \delta^a{}_c \delta^d{}_b) \delta (S^{\pm 1})^c{}{}_d.
    \end{split}
\end{equation}

Equations~\eqref{eqn:e_prop} and~\eqref{eqn:Y_prop} imply that
\begin{equation}
    \mc V^{ab} = \frac{\Lambda_g}{m^2} g^{ab}, \qquad \tilde{\mc V}^{ab} = \frac{\Lambda_f}{m^2 \alpha^2 c^4} g^{ab},
\end{equation}
where
\begin{subequations} \label{eqn:prop_Lambda}
    \begin{align}
        \Lambda_g &= m^2 (\beta_0 + 3 c \beta_1 + 3 c^2 \beta_2 + c^3 \beta_3), \\
        \Lambda_f &= \frac{m^2}{\alpha^2 c^2} (c \beta_1 + 3 c^2 \beta_2 + 3 c^3 \beta_3 + c^4 \beta_4).
    \end{align}
\end{subequations}
As such, the background Einstein equations in Eq.~\eqref{eq:FieldEquations} become
\begin{equation} \label{eqn:prop_einstein}
    G^{ab} + \Lambda_g g^{ab} = 0 = G^{ab} + \Lambda_f g^{ab}.
\end{equation}
This set of equations is only consistent if $\Lambda_g = \Lambda_f$, which provides a constraint in terms of a quartic polynomial in $c$ (given the coefficients $\beta_n$).
Moreover, since this background has a non-vanishing cosmological constant, it cannot be asymptotically flat, unless $\Lambda = 0$.
For arbitrary values of the coefficients $\beta_n$, this cannot simultaneously be enforced while also demanding that $\Lambda_g = \Lambda_f$ (as this would require simultaneously solving the polynomial equations obtained by setting the right-hand sides of Eq.~\eqref{eqn:prop_Lambda} to zero with a single $c$).
As such, one can think of this as a constraint on the coefficients $\beta_n$.

The real advantage of the proportional background, however, is in simplifying the formula for $\delta S^a{}_b$, which takes the form
\begin{equation}
    \delta S^a{}_b = \frac{1}{2c} g^{ac} (\delta f_{cb} - c^2 \delta g_{cb}),
\end{equation}
which comes from expanding the variation of Eq.~\eqref{eq:SDef}.
Moreover, using Eqs.~\eqref{eq:UtensorSymPoly} and~\eqref{eqn:var_Y}, together with the fact that $\delta (S^{-1})^a{}_b = -\delta S^a{}_b$, one finds that
\begin{equation}
    \ord{1}{\mc U}^a{}_{bc}{}^d \delta S^c{}_d = -c^4 \ord{1}{\tilde{\mc U}}^a{}_{bc}{}^d \delta S^c{}_d = \frac{m_{\rm FP}^2}{m^2} \frac{\alpha^2 c}{1 + \alpha^2 c^2} (\delta^a{}_b \delta S^c{}_c - \delta S^a{}_b),
\end{equation}
where
\begin{equation}
    m_{\rm FP}^2 \equiv m^2 \left(1 + \frac{1}{\alpha^2 c^2}\right) (c \beta_1 + 2c^2 \beta_2 + c^3 \beta_3).
\end{equation}
Note that, like $m$, $m_{\rm FP}$ has units of inverse length.
As such, using Eq.~\eqref{eqn:var_V}, Eq.~\eqref{eq:FieldEquations_Linearized} becomes
\begin{subequations}
    \begin{align}
        \ord{1} G^{ab} \{\delta \bs g\} &= \Lambda g^{ac} g^{bd} \delta g_{cd} - m_{\rm FP}^2 \frac{\alpha^2 c}{1 + \alpha^2 c^2} (g^{ab} \delta S^c{}_c - g^{bc} \delta S^a{}_c), \label{eq:FE1} \\
        \ord{1} G^{ab} \{\delta \bs f\} &= \Lambda g^{ac} g^{bd} \delta f_{cd} + m_{\rm FP}^2 \frac{c}{1 + \alpha^2 c^2} (g^{ab} \delta S^c{}_c - g^{bc} \delta S^a{}_c) \label{eq:FE2}.
    \end{align}
\end{subequations}

Following the work of~\cite{Bernard:2015uic}, this form of the linearized Einstein equations can be decoupled by defining
\begin{equation}
    \gamma_{ab} (\lambda) \equiv g_{ab} (\lambda) + \alpha^2 f_{ab} (\lambda), \qquad \phi_{ab} (\lambda) \equiv g_{ac} S^c{}_b (\lambda),
\end{equation}
such that
\begin{equation}
  \label{eq:delta_gamma_phi_def}
    \delta \gamma_{ab} = \delta g_{ab} + \alpha^2 \delta f_{ab}, \qquad \delta \phi_{ab} = \frac{1}{2c} (\delta f_{ab} - c^2 \delta g_{ab}).
\end{equation}
Adding together Eq.~\eqref{eq:FE1} and $\alpha^2$ times Eq.~\eqref{eq:FE2} shows that $\delta \gamma_{ab}$ obeys the linearized Einstein equations with cosmological constant $\Lambda$:
\begin{equation} \label{eqn:eigenstate_eom1}
    \ord{1} G^{ab} \{\delta \bs \gamma\} = \Lambda g^{ac} g^{bd} \delta \gamma_{cd}.
\end{equation}
On the contrary, subtracting $c^2$ times Eq.~\eqref{eq:FE1} from Eq.~\eqref{eq:FE2}, and then dividing by an overall factor of $2c$ shows that $\delta \phi_{ab}$ obeys the massive Fierz-Pauli equations, with mass $m_{\rm FP}$:
\begin{equation} \label{eqn:eigenstate_eom2}
    \ord{1} G^{ab} \{\delta \bs \phi\} = \left[\left(\Lambda - \frac{m_{\rm FP}^2}{2}\right) g^{ac} g^{bd} + \frac{m_{\rm FP}^2}{2} g^{ab} g^{cd}\right] \delta \phi_{cd}.
\end{equation}

An interesting property of $\delta \phi_{cd}$ is that it is gauge-invariant: under a linearized diffeomorphism, since
\begin{equation}
    \delta g_{ab} \to \delta g_{ab} + \lie_\xi g_{ab}, \qquad \delta f_{ab} \to \delta f_{ab} + c^2 \lie_\xi g_{ab},
\end{equation}
it follows that
\begin{equation}
    \delta \gamma_{ab} \to \delta \gamma_{ab} + (1 + \alpha^2 c^2) \lie_\xi g_{ab}, \qquad \delta \phi_{ab} \to \delta \phi_{ab}.
\end{equation}
Moreover, by taking the divergence and trace of Eq.~\eqref{eqn:eigenstate_eom2}, one can show that (see, for example,~\cite{Deffayet2011})
\begin{equation}
    g^{ab} \nabla_a \delta \phi_{bc} = 0, \qquad g^{ab} \delta \phi_{ab} = 0.
\end{equation}

\section{Conserved quantities}
\label{sec:currents}

We now consider the problem of finding a suitable notion of an energy flux in bigravity.
For simplicity, we only consider the problem of defining such a quantity for perturbative solutions: that is, we only want to construct a conserved current out of $g_{ab}$, $f_{ab}$, $\delta g_{ab}$, $\delta f_{ab}$, and a Killing vector $\xi^a$.

In this section, we first consider the most commonly adopted approach to conserved currents in linearized gravity, the Isaacson stress-energy tensor, and discuss the various issues which arise in extending it to bigravity.
We then consider the (much simpler) canonical current, and show that it is equivalent to a current, which we call the ``effective source current,'' generalizing the conserved current constructed from the effective stress-energy tensor of general relativity.
We then apply this general framework to the case of bigravity, both generally and in the case where the backgrounds $g_{ab}$ and $f_{ab}$ are proportional.
Finally, we discuss a different approach to conserved quantities in bigravity through asymptotic charges (defined, for example, through the Wald-Zoupas procedure~\cite{Wald:1999wa}), and why we do not consider these charges in this paper.

\subsection{Difficulties with the Isaacson stress-energy tensor} \label{sec:isaacson}

\figLengths

One way that has often been used to find such currents is to start with a notion of an energy momentum tensor through the ``Isaacson'' approach~\cite{Isaacson:1968hbi, Isaacson:1968zza}.
In source-free GR, the use of the Einstein equations at first order gives the equations of motion, while at the second order the source term for the Einstein equations is an ``effective source'' given by the second-order perturbation to the Einstein tensor:
\begin{equation}
    \ord{1} G^{ab} \{\delta^2 \bs g\} = -\frac{1}{2} \ord{2} G^{ab} \{\delta \bs g, \delta \bs g\}
\end{equation}
(the operator on the right-hand side of this equation is defined explicitly below).
Moreover, assuming that the perturbations $\delta g_{ab}$ are high-frequency solutions, it becomes appropriate to average the right-hand side of this equation, writing
\begin{equation} \label{eqn:isaacson}
    T^{ab}_I \equiv -\frac{1}{16\pi} \avg{\ord{2} G^{ab} \left\{\delta \bs g, \delta \bs g\right\}},
\end{equation}
where $\avg{\cdots}$ denotes a short-wavelength averaging procedure, for example the Brill-Hartle approach~\cite{Brill:1964zz}, or as described by Zalaletdinov~\cite{Zalaletdinov}.
It is this averaging procedure which becomes ambiguous in the extension to bigravity.

As understood in GR, an averaging procedure is used in the presence of small perturbations on a strongly curved background~\cite{Brill:1964zz}.
Waves are taken to be small enough on the background to be analyzed in a linearized gravity approach, where their wavelengths $\lambda_{\rm GW}$ are short compared to the curvature scale $L_{\rm ST}$ of the spacetime on which they propagate: $L_{\rm ST} \gg \lambda_{\rm GW}$.

For GWs in vacuum GR on a curved background, there are only two length scales to consider.
However, in bigravity, we have additional length scales to consider, as represented in Fig.~\ref{fig:Wavelength}.  First, the procedure is already ambiguous because there are two metrics.  The Brill-Hartle approach requires using the parallel propagator of one metric---does it matter which one?

Second, the massive degree of freedom introduces its own length scale, the Compton wavelength of the graviton, $\lambda_c$.
In GR, when there are only two length scales $\lambda_{\rm GW} \ll L_{\rm ST}$ which are parametrically separated, we only need to choose an arbitrary averaging length scale $L_{\rm avg}$ which separates the two, $\lambda_{\rm GW} \ll L_{\rm avg} \ll L_{\rm ST}$.  With the introduction of $\lambda_c$, we need to know the hierarchy between $\lambda_c$ and $L_{\rm ST}$, and whether $L_{\rm avg}$ will end up averaging over $\lambda_c$.  If the averaging neglects mass terms, then the results would effectively take the massless limit, thus neglecting a fundamental property of the graviton.

We should also recall that one of the original motivations for dRGT and bigravity were cosmological in nature, to make the gravity theory self-accelerate.  So, we should also have a cosmological length scale $L_{\rm cosm}\sim cH_0^{-1}$.  Whether the curvature radius $L_{\rm ST}$ is equal to $L_{\rm cosm}$ depends on how close we are to the source---if the curvature is dominated by the Coulombic contribution $\sim GM/R^3$, or by the Hubble contribution $1/L_{\rm cosm}^2$.  In the latter case, and assuming that dark energy is due to bigravity, $L_{\rm cosm}$ is related to the graviton mass and thus $\lambda_c$, but also the coupling constants and dynamical solution, as seen in Eq.~\eqref{eqn:prop_Lambda}.

Finally, there is the question of which effective stress-energy tensor gives ``the'' energy.  In GR, there is only one metric equation of motion, so averaging the second variation of that equation generates an effective stress-energy tensor.  In bigravity, we have two metric equations, one each for $g_{ab}$ and $f_{ab}$.  If we followed the Isaacson procedure, we would have an effective stress-energy tensor on the right-hand side of each equation.  How would we interpret the two stress tensors?  Is the energy loss present in both tensors, or do we need to take some appropriate combination of the two?

Given the variety of scales to consider in this theory, there is no clear approach as to the ``proper'' way to apply an averaging scheme similar to the one utilized by Isaacson.
Even if one were to make particular assumptions about the length scales to ensure a straightforward mathematical approach, the meaning of the resulting quantity will be unclear.

As such, one is motivated to consider neglecting the averaging part of the Isaacson approach, instead simply defining the right-hand side of Eq.~\eqref{eqn:isaacson}, \emph{without} the average, as the effective stress-energy tensor.
However, this reveals another issue: this tensor is no longer conserved in bigravity: while $\nabla_a G^{ab} = 0$, one has that
\begin{equation}
  \nabla_a \ord{2} G^{ab} \{\delta \bs g, \delta \bs g\} \neq 0.
\end{equation}
The vanishing divergence in GR came from $\ord{1} G^{ab} \{\delta \bs g\} = 0$.
Instead, the analogue of the effective stress-energy tensor that will arise in bigravity should be some combination of $\ord{2} G^{ab} \{\delta \bs g, \delta \bs g\}$, the corresponding object constructed from $f_{ab}$ and $\delta f_{ab}$, along with terms related to the variations of the source terms in the equations of motion, $\mc V^{ab}$ and $\tilde{\mc V}^{ab}$.
Very quickly, the equations that arise become quite complicated; a far simpler current, which we consider in the next section, is the canonical current.

\subsection{Canonical current} \label{sec:canonical}

In this section, we define a conserved current, which we refer to (following~\cite{Bonga2019}) as the \emph{canonical current}~\cite{Lee1990, Burnett1990}, and consider its properties in a variety of field theories.
This current is directly constructed from the Lagrangian of any field theory, and has a wide array of applications: its integral over a Cauchy surface (the \emph{canonical energy}) appears in the analysis of the stability of black hole spacetimes~\cite{Hollands2012, Prabhu2015a} and rotating relativistic stars~\cite{Friedman1978, Prabhu2016}, as well as the proof that black holes cannot by overcharged or overspun~\cite{Sorce2017}.
Moreover, the methods used to construct this current have proven useful for defining additional types of conserved currents, not related to isometries of a spacetime, that may provide methods to understand the evolution of point particles under the gravitational self-force~\cite{Grant2019, Grant2020, Grant2022}.

\subsubsection{General formalism}

We start by considering a general, diffeomorphism-invariant theory without internal symmetries.\footnote{For the case where there are internal symmetries, see the discussion in~\cite{Prabhu2015b}.}
For these theories, a particular type of variation is given by those with respect to the action of diffeomorphisms characterized by some vector field $\xi^a$.
This variation $\delta_\xi$, acting on any tensor field $\bs Q$, is given by the Lie derivative:
\begin{equation} \label{eqn:lie_var}
  \delta_\xi \bs Q = \lie_\xi \bs Q.
\end{equation}
In particular, for the Lagrangian four-form $\bs L$, we have that
\begin{equation}
  \delta_\xi \bs L = \ud (\bs \xi \cdot \bs L),
\end{equation}
which follows from Cartan's magic formula for differential forms,
\begin{equation}
  \lie_\xi \bs \omega = \bs \xi \cdot (\ud \bs \omega) + \ud (\bs \xi \cdot \bs \omega),
\end{equation}
and the fact that, as a four-form, $\ud \bs L = 0$.
Note that this is a special case of the variation of $\bs L$ under a symmetry: under a symmetry, the Lagrangian four-form \emph{must} transform by a boundary term in order to preserve the equations of motion.

Next, consider an arbitrary variation of the Lagrangian.
Denote the dynamical fields of the theory by $\Phi_A$, where $A$ is an abstract index in a vector space in which these fields live; for the example of bigravity, for example, we write
\begin{equation}
  \Phi_A \equiv \begin{pmatrix}
    g_{ab}, & f_{ab}
  \end{pmatrix},
\end{equation}
where $g_{ab}$ and $f_{ab}$ are the two metrics, and so $A$ is an index on the Cartesian product of two copies of the space of symmetric, rank two tensor fields.
By the product rule, one can always write
\begin{equation} \label{eqn:var_L}
  \delta \bs L \equiv \bs E^A \delta \Phi_A + \ud (\bs \theta\{\delta \bs \Phi\}),
\end{equation}
where we suppress the differential form indices on $\bs E^A$, which gives the \emph{equations of motion} of the theory, and where $\bs \theta$ is a linear functional of $\delta \Phi_A$ (which we indicate with the curly braces) called the \emph{presymplectic potential}.

The presymplectic potential can be used to generate two types of conserved current, both of which are relevant to this discussion.
The first is the \emph{Noether current}~\cite{Lee1990, Iyer1994}, given by
\begin{equation}
  \bs J_\xi \equiv \bs \theta\{\lie_\xi \bs \Phi\} - \bs \xi \cdot \bs L.
\end{equation}
This current is conserved when the equations of motion of the theory hold:
\begin{equation} \label{eqn:dJ}
  \ud \bs J_\xi = \ud \bs \theta \{\lie_\xi \bs \Phi\} - \ud (\bs \xi \cdot \bs L) = -\bs E^A \lie_\xi \Phi_A.
\end{equation}
Since $\lie_\xi$ involves taking derivatives of $\xi^a$, one can use the product rule on the right-hand to yield~\cite{Iyer1994, Seifert2006}
\begin{equation} \label{eqn:constraint_def}
  \ud \bs J_\xi \equiv \ud (\xi^a \bs{\mc C}_{aB} \cdot \bs E^B) + \xi^a \bs U_a \{\bs E\},
\end{equation}
where we drop the vector field indices on $\bs{\mc C}_{aB}$ that are contracted into the differential form indices on $\bs E^B$ (giving a three-form).
The fact that the boundary term is a linear function (and not \emph{functional}) of the equations of motion $\bs E^A$ follows from the fact that the right-hand side of Eq.~\eqref{eqn:dJ} is also a linear function of $\bs E^A$.
Moreover, the boundary term is a linear function of $\xi^a$, due to the fact that Lie derivatives involve at most a single derivative of $\xi^a$.

Equation~\eqref{eqn:constraint_def} can be used to construct the so-called \emph{Noether charge}~\cite{Lee1990, Iyer1994}, which follows from the fact that $\bs U_a \{\bs E\} = 0$, which we now prove, following~\cite{Seifert2006}.
If $\bs U_a \{\bs E\} \neq 0$, then there is some $\xi^a$ with support in some compact region $V$ such that, for a given volume for $\bs \epsilon$, $\xi^a \bs U_a \{\bs E\} = \alpha \bs \epsilon$, with $\alpha \geq 0$ (but $\alpha$ not everywhere $0$).
Rearranging Eq.~\eqref{eqn:constraint_def} and integrating over $\partial V$, we have that
\begin{equation}
  \int_{\partial V} (\bs J_\xi - \xi^a \bs{\mc C}_{aB} \cdot \bs E^B) = \int_V \xi^a \bs U_a \{\bs E\} > 0.
\end{equation}
However, since $\xi^a$ only has support in $V$ and $V$ is compact, it follows that the left-hand side must also be zero.
This is a contradiction, and so
\begin{equation}
  \bs U_a \{\bs E\} = 0.
\end{equation}
In the case of general relativity, this equation implies that $\nabla_b G^{ab} = 0$.
For arbitrary theories, we have that
\begin{equation}
  \ud (\bs J_\xi - \xi^a \bs{\mc C}_{aB} \cdot \bs E^B) = 0.
\end{equation}
At this point, we can apply Wald's theorem~\cite{Wald1990}, which yields
\begin{equation} \label{eqn:Q_def}
  \bs J_\xi = \xi^a \bs{\mc C}_{aB} \cdot \bs E^B + \ud \bs Q_\xi,
\end{equation}
where the two-form $\bs Q_\xi$ is the \emph{Noether charge}.
In particular, when the equations of motion hold, $\bs J_\xi$ is a total derivative.
The quantity $\bs{\mc C}_{aB} \cdot \bs E^B$ is known as the \emph{constraint} of the theory.\footnote{See Eq.~(2.35) of~\cite{Prabhu2015b} for an expression for the constraint that is also explicitly linear in $\bs E$, in a different class of theories than those considered here.}

The next current that can be constructed from the presymplectic potential is the \emph{symplectic current}~\cite{Lee1990, Burnett1990}:
\begin{equation} \label{eqn:omega_def}
  \bs \omega\{\delta_1 \bs \Phi, \delta_2 \bs \Phi\} \equiv \delta_1 \bs \theta\{\delta_2 \bs \Phi\} - \delta_2 \bs \theta\{\delta_1 \bs \Phi\},
\end{equation}
which is an antisymmetric, bilinear functional of two variations $\delta_1 \Phi_A$ and $\delta_2 \Phi_A$, assuming that $\delta_1$ and $\delta_2$ commute (otherwise there would be a dependence on $[\delta_2, \delta_1] \Phi_A$).
Whenever $\delta_1 \Phi_A$ and $\delta_2 \Phi_A$ satisfy the linearized equations of motion, $\bs \omega$ is also a conserved current, as
\begin{equation}
  \begin{split}
    \ud \bs \omega\{\delta_1 \bs \Phi, \delta_2 \bs \Phi\} &= \delta_1 \ud \bs \theta\{\delta_2 \bs \Phi\} - \delta_2 \ud \bs \theta\{\delta_1 \bs \Phi\} \\
    &= \delta_2 \bs E^A \delta_1 \Phi_A - \delta_1 \bs E^A \delta_2 \Phi_A,
  \end{split}
\end{equation}
which uses the fact that $\delta_1$ and $\delta_2$ commute, together with Eq.~\eqref{eqn:var_L} for the variation of the Lagrangian.
Integrating $\bs \omega$ over a Cauchy surface provides a notion of a symplectic product on phase space for the field~\cite{Lee1990}.

Moreover, Eq.~\eqref{eqn:Q_def} can be used to show that the symplectic current is gauge-invariant (that is, invariant under linearized diffeomorphisms), up to a boundary term.
Under a change of gauge, $\delta \Phi_A \to \delta \Phi_A + \lie_\xi \Phi_A$ (for $\delta = \delta_1$ or $\delta_2$), and so the bilinearity of $\bs \omega$ implies that we can reduce this problem to considering
\begin{equation}
  \bs \omega\{\delta \bs \Phi, \lie_\xi \bs \Phi\} = \delta \bs \theta\{\lie_\xi \bs \Phi\} - \lie_\xi \bs \theta\{\delta \bs \Phi\},
\end{equation}
where we have used the fact that $\lie_\xi = \delta_\xi$; note that $[\delta, \delta_\xi] = 0$.
Using the definition of the Noether current for the first term, together with Cartan's magic formula for the second, yields
\begin{equation} \label{eqn:omega_xi}
  \begin{split}
    \bs \omega\{\delta \bs \Phi, \lie_\xi \bs \Phi\} &= \delta \bs J_\xi + \bs \xi \cdot (\delta \bs L - \ud \bs \theta\{\delta \bs \Phi\}) - \ud (\bs \xi \cdot \bs \theta\{\delta \bs \Phi\}) \\
    &= \bs \xi \cdot (\bs E^A \delta \Phi_A) + \xi^a \delta (\bs{\mc C}_{aB} \cdot \bs E^B) + \ud (\delta \bs Q_\xi - \bs \xi \cdot \bs \theta\{\delta \bs \Phi\}),
  \end{split}
\end{equation}
where we have used the variation of the Lagrangian and Eq.~\eqref{eqn:Q_def}.
As such, when $\bs E = 0$ and $\delta \bs E = 0$, we find that
\begin{equation}
  \bs \omega\{\delta \bs \Phi, \lie_\xi \bs \Phi\} = \ud (\delta \bs Q_\xi - \bs \xi \cdot \bs \theta\{\delta \bs \Phi\}),
\end{equation}
which is a boundary term.

Using the symplectic current, we can define the canonical current by
\begin{equation} \label{eqn:canonical}
  \bs{\mc E}_\xi \{\delta_1 \bs \Phi, \delta_2 \bs \Phi\} \equiv \bs \omega\{\delta_1 \bs \Phi, \lie_\xi \delta_2 \bs \Phi\},
\end{equation}
where $\delta_1$ and $\delta_2$ commute.
Now, take a variation of Eq.~\eqref{eqn:omega_xi}, which gives
\begin{equation} \label{eqn:var_omega}
  \delta_2 \bs \omega\{\delta_1 \bs \Phi, \lie_\xi \bs \Phi\} = \bs \xi \cdot \delta_2 \left(\bs E^A \delta_1 \bs \Phi_A\right) + \xi^a \delta_1 \delta_2 (\bs{\mc C}_{aB} \cdot \bs E^B) + \ud \delta_2 (\delta_1 \bs Q_\xi - \bs \xi \cdot \bs \theta\{\delta_1 \bs \Phi\}).
\end{equation}
This equation can be used to relate the canonical current to the effective stress-energy tensor.
When we perform an independent variation $\delta_1$ of a multilinear functional $\ord{m}{\bs Q} \{\delta_1 \bs \Phi, \ldots, \delta_n \bs \Phi\}$, there are two contributions.  The first is from the variation of its (possibly nonlinear) dependence on the background fields, and the second is the higher variations of all the arguments.  This allows us to generalize the notation introduced in Eq.~\eqref{eqn:ord_1} and so define $\ord{m + 1}{\bs Q}$ via
\begin{equation} \label{eqn:ord_n1}
    \delta_1 \ord{m}{\bs Q} \{\delta_2 \bs \Phi, \ldots, \delta_{n + 1} \bs \Phi\} \equiv \ord{m + 1}{\bs Q} \{\delta_1 \bs \Phi, \ldots, \delta_{n + 1} \bs \Phi\} + \sum_{i = 1}^n \ord{m}{\bs Q} \{\delta_2 \bs \Phi, \ldots, \delta_1 \delta_{i + 1} \bs \Phi, \ldots, \delta_{n + 1} \bs \Phi\}.
\end{equation}
As an example, $\ord{2}{\bs E}^A \{\delta_1 \bs \Phi, \delta_2 \bs \Phi\}$ is given by
\begin{equation} \label{eqn:ord_2_E}
  \delta_1 \delta_2 \bs E^A =
  \delta_1 \ord{1}{\bs E}^A \{ \delta_2 \bs \Phi \} =
  \ord{2}{\bs E}^A \{\delta_1 \bs \Phi, \delta_2 \bs \Phi\} + \ord{1}{\bs E}^A \{\delta_1 \delta_2 \bs \Phi\}.
\end{equation}
Note that, in Eq.~\eqref{eqn:ord_n1}, $m$ does not necessarily equal $n$.
By convention, in the case where the order $m$ is absent, it is taken to be zero; an example of such a case is the definition of $\ord{1}{\bs \omega} \{\delta_1 \bs \Phi, \delta_2 \bs \Phi, \delta_3 \bs \Phi\}$:
\begin{equation}
  \delta_1 \bs \omega \{\delta_2 \bs \Phi, \delta_3 \bs \Phi\} \equiv \ord{1}{\bs \omega} \{\delta_1 \bs \Phi, \delta_2 \bs \Phi, \delta_3 \bs \Phi\} + \bs \omega \{\delta_1 \delta_2 \bs \Phi, \delta_3 \bs \Phi\} + \bs \omega \{\delta_2 \bs \Phi, \delta_1 \delta_3 \bs \Phi\}.
\end{equation}

Using this new notation, the canonical current is related to variations of the symplectic current by
\begin{equation} \label{eqn:canonical_var_omega}
  \bs{\mc E}_\xi \{\delta_1 \bs \Phi, \delta_2 \bs \Phi\} = \delta_2 \bs \omega\{\delta_1 \bs \Phi, \lie_\xi \bs \Phi\} - \ord{1}{\bs \omega} \{\delta_2 \bs \Phi, \delta_1 \bs \Phi, \lie_\xi \bs \Phi\} - \bs \omega\{\delta_1 \delta_2 \bs \Phi, \lie_\xi \bs \Phi\}.
\end{equation}
Similarly, note that
\begin{equation} \label{eqn:var2_constraint}
  \delta_1 \delta_2 (\bs{\mc C}_{aB} \cdot \bs E^B) = \delta_1 \delta_2 \bs{\mc C}_{aB} \cdot \bs E^B + \delta_1 \bs{\mc C}_{aB} \cdot \delta_2 \bs E^B + \delta_2 \bs{\mc C}_{aB} \cdot \delta_1 \bs E^B + \bs{\mc C}_{aB} \cdot \delta_1 \delta_2 \bs E^B,
\end{equation}
When the second-order equations of motion are imposed, note that Eq.~\eqref{eqn:ord_2_E} implies that $-\ord{2}{\bs E}^A \{\delta_1 \bs \Phi, \delta_2 \bs \Phi\}$ is the source for the differential equation satisfied by the second-order perturbation.

It is from $\ord{2}{\bs E}^A \{\delta_1 \bs \Phi, \delta_2 \bs \Phi\}$ that we will define the effective source current.
To do so, note that these equations hold, regardless of the vanishing of the equations of motion, and so we are free to set $\delta_1 \delta_2 \Phi_A = 0$, and impose only the zeroth and first-order equations of motion.
Equations~\eqref{eqn:var_omega}, \eqref{eqn:canonical_var_omega}, \eqref{eqn:var2_constraint}, and~\eqref{eqn:ord_2_E} then give
\begin{equation}
  \bs{\mc E}_\xi \{\delta_1 \bs \Phi, \delta_2 \bs \Phi\} = \xi^a \bs{\mc C}_{aB} \cdot \ord{2}{\bs E}^B \{\delta_1 \bs \Phi, \delta_2 \bs \Phi\} + \ud \delta_2 (\delta_1 \bs Q_\xi - \bs \xi \cdot \bs \theta\{\delta_1 \bs \Phi\}) - \ord{1}{\bs \omega} \{\delta_2 \bs \Phi, \delta_1 \bs \Phi, \lie_\xi \bs \Phi\}.
\end{equation}
In the case where $\xi^a$ is a background symmetry, so that $\lie_\xi \bs \Phi = 0$, it follows that the canonical current is given by:
\begin{equation}
  \begin{split}
    \bs{\mc E}_\xi \{\delta_1 \bs \Phi, \delta_2 \bs \Phi\} &= \xi^a \bs{\mc C}_{aB} \cdot \ord{2}{\bs E}^B \{\delta_1 \bs \Phi, \delta_2 \bs \Phi\} - \ud \delta_2 (\delta_1 \bs Q_\xi + \bs \xi \cdot \bs \theta\{\delta_1 \bs \Phi\}) \\
    &\equiv \bs{\mc S}_\xi \{\delta_1 \bs \Phi, \delta_2 \bs \Phi\} - \ud \delta_2 (\delta_1 \bs Q_\xi - \bs \xi \cdot \bs \theta\{\delta_1 \bs \Phi\}).
  \end{split}
\end{equation}
The effective source current $\bs{\mc S}_\xi$, in the case of general relativity, when applied to two equal variations $\delta_1 g_{ab} = \delta_2 g_{ab} = \delta g_{ab}$, is related to the conserved current arising from the effective stress-energy tensor.
Moreover, this current, like the canonical current, must be conserved in the case where the linearized equations of motion hold, and is gauge-invariant up to a boundary term.

One final issue to note is that there are two potential ambiguities in the canonical current.
First, the Lagrangian $\bs L$ is only defined up to a boundary term:
\begin{equation}
    \bs L \to \bs L + \ud \bs \eta
\end{equation}
will yield the same equations of motion, modifying the presymplectic potential by
\begin{equation}
    \bs \theta\{\delta \bs \Phi\} \to \bs \theta\{\delta \bs \Phi\} + \delta \bs \eta.
\end{equation}
Note, however, that this is the addition of a total variation, and so does not affect the symplectic current, and therefore the canonical current.
On the other hand, the presymplectic potential itself has an ambiguity of the form
\begin{equation}
    \bs \theta\{\delta \bs \Phi\} \to \bs \theta\{\delta \bs \Phi\} + \ud \bs \Upsilon\{\delta \bs \Phi\},
\end{equation}
since it is only defined in terms of its exterior derivative.
This yields a modification of the symplectic and canonical currents by boundary terms, which in the latter case is given by 
\begin{equation}
    \bs{\mc E}_\xi \{\delta_1 \bs \Phi, \delta_2 \bs \Phi\} \to \bs{\mc E}_\xi \{\delta_1 \bs \Phi, \delta_2 \bs \Phi\} + \ud \left(\ord{1}{\bs \Upsilon} \{\delta_1 \bs \Phi, \lie_\xi \delta_2 \bs \Phi\} - \ord{1}{\bs \Upsilon} \{\lie_\xi \delta_2 \bs \Phi, \delta_1 \bs \Phi\}\right).
\end{equation}

\subsubsection{Specialization to bigravity} \label{sec:bigravity_currents}

We now apply this general formalism to the case of bigravity.
To vary the Lagrangian in Eq.~\eqref{eq:BigravLagrangian}, we use the fact that
\begin{equation} \label{eqn:var_epsilon}
  \delta \bs \epsilon = \frac{1}{2} \delta g_{ab} g^{ab} \bs \epsilon,
\end{equation}
with the same expression holding for $\tilde{\bs \epsilon}$, replacing $\delta g_{ab}$ with $\delta f_{ab}$ and $g^{ab}$ with $f^{ab}$.
Now, denoting by $C^a{}_{bc} (\lambda)$ the connection coefficient between $\nabla_a (\lambda)$ and $\nabla_a$ (that is,
\begin{equation}
    [\nabla_a (\lambda) - \nabla_a] v^b = C^b{}_{ac} (\lambda) v^c,
\end{equation}
for any vector $v^a$), we find that
\begin{equation} \label{eqn:var_R}
  g^{ab} \delta R_{ab} = 2 g^{ab} \nabla_{[c} \delta C^c{}_{a]b} = \nabla_a v^a \{\delta \bs g\},
\end{equation}
where
\begin{equation}
  v^a \{\delta \bs g\} \equiv 2 \delta C^{[a}{}_{bc} g^{b]c}.
\end{equation}
A similar expression holds for $f^{ab} \delta \tilde R_{ab}$:
\begin{equation} \label{eqn:var_tildeR}
  f^{ab} \delta \tilde R_{ab} = \tilde \nabla_a \tilde v^a \{\delta {\bs f}\}.
\end{equation}

Combining Eq.~\eqref{eqn:var_epsilon}, \eqref{eqn:var_R}, \eqref{eqn:var_tildeR}, and~\eqref{eq:Vtensors}, we find that
\begin{equation}
  \delta \bs L = -\left[(G^{ab} + m^2 \mc V^{ab}) \delta g_{ab} \bs \epsilon + (\alpha^2 \tilde G^{ab} + m^2 \tilde{\mc V}^{ab}) \delta f_{ab} \tilde{\bs \epsilon}\right] + \nabla_a v^a \{\delta \bs g\} \bs \epsilon + \alpha^2 \tilde \nabla_a \tilde v^a \{\delta {\bs f}\} \tilde{\bs \epsilon}.
\end{equation}
To get this in the form of Eq.~\eqref{eqn:var_L}, we see that
\begin{equation} \label{eqn:big_E}
  \bs E^A \equiv -\begin{pmatrix}
    (G^{ab} + m^2 \mc V^{ab}) \bs \epsilon \\
    (\alpha^2 \tilde G^{ab} + m^2 \tilde{\mc V}^{ab}) \tilde{\bs \epsilon}
  \end{pmatrix},
\end{equation}
and, using the fact that \{Eq.~(B.2.22) of Wald~\cite{Wald1984}\}
\begin{equation}
  (\nabla_e v^e) \epsilon_{abcd} = \ud (\bs v \cdot \bs \epsilon),
\end{equation}
we have that
\begin{equation} \label{eqn:big_theta}
  \bs \theta\{\delta \bs \Phi\} = \bs v \{\delta \bs g\} \cdot \bs \epsilon + \alpha^2 \tilde{\bs v} \{\delta {\bs f}\} \cdot \tilde{\bs \epsilon}.
\end{equation}

Next, we need to find the constraint.
This can be done by noting that
\begin{equation}
  -\bs E^A \lie_\xi \Phi_A = 2 (G^{ab} + m^2 \mc V^{ab}) g_{bc} (\nabla_a \xi^c) \bs \epsilon + 2 (\alpha^2 \tilde G^{ab} + m^2 \tilde{\mc V}^{ab}) f_{bc} (\tilde \nabla_a \xi^c) \tilde{\bs \epsilon}.
\end{equation}
As such, using Eq.~\eqref{eqn:constraint_def},
\begin{equation}
  (\mc C_{dE} \cdot E^E)_{abc} = 2 [(G^{ef} + m^2 \mc V^{ef}) g_{fd} \epsilon_{eabc} + (\alpha^2 \tilde G^{ef} + m^2 \tilde{\mc V}^{ef}) f_{fd} \tilde \epsilon_{eabc}],
\end{equation}
which implies that, writing $\mc C_{aB}{}^c$ as a row vector,
\begin{equation} \label{eqn:big_C}
  \mc C_{aB}{}^d = -2 \begin{pmatrix}
    g_{a(b} \delta^d{}_{c)}, & f_{a(b} \delta^d{}_{c)}
  \end{pmatrix}.
\end{equation}
Similarly, we have that
\begin{equation}
    \bs U_a \{\bs E\} = 2 g_{ac} [\nabla_b (G^{bc} + m^2 \mc V^{bc})] \bs \epsilon + 2 f_{ac} [\tilde \nabla_b (\alpha^2 \tilde G^{bc} + m^2 \tilde{\mc V}^{bc})] \tilde{\bs \epsilon} = 0,
\end{equation}
from which Eq.~\eqref{eqn:div_identity} follows by the Bianchi identities for $G^{ab}$ and $\tilde G^{ab}$.

Using Eq.~\eqref{eqn:big_theta}, we can write down the symplectic current, and therefore the canonical current.
First, using Eq.~\eqref{eqn:ord_n1} we can define $\ord{1} v^a \{\delta_1 \bs g, \delta_2 \bs g\}$, which is given by
\begin{equation}
  \ord{1} v^a \{\delta_1 \bs g, \delta_2 \bs g\} = 2 \delta_2 C^{[a}{}_{bc} \delta_1 g^{b]c}.
\end{equation}
A similar expression gives $\ord{1}{\tilde v}^a \{\delta_1 \tilde{\bs g}, \delta_2 \tilde{\bs g}\}$.
For the symplectic current, we therefore find that
\begin{equation} \label{eqn:big_omega}
  \bs \omega\{\delta_1 \bs \Phi, \delta_2 \bs \Phi\} = \bs w \{\delta_1 \bs g, \delta_2 \bs g\} \cdot \bs \epsilon + \alpha^2 \tilde{\bs w} \{\delta_1 {\bs f}, \delta_2 {\bs f}\} \cdot \tilde{\bs \epsilon},
\end{equation}
where
\begin{equation}
  w^a \{\delta_1 \bs g, \delta_2 \bs g\} \equiv \ord{1} v^a \{\delta_1 \bs g, \delta_2 \bs g\} + \frac{1}{2} g^{bc} \delta_1 g_{bc} v^a \{\delta_2 \bs g\} - (\interchange{1}{2}),
\end{equation}
with a similar expression holding for $\tilde w^a$.
To use the same notation as~\cite{Burnett1990},
\begin{equation} \label{eqn:w}
  w^a \{\delta_1 \bs g, \delta_2 \bs g\} = P^{abcdef} \left[\delta_1 g_{bc} \nabla_d \delta_2 g_{ef} - (\interchange{1}{2})\right],
\end{equation}
where
\begin{equation}
  P^{abcdef} \equiv g^{ae} g^{fb} g^{cd} - \frac{1}{2} g^{ad} g^{be} g^{fc} - \frac{1}{2} g^{ab} g^{cd} g^{ef} - \frac{1}{2} g^{bc} g^{ae} g^{fd} + \frac{1}{2} g^{bc} g^{ad} g^{ef}.
\end{equation}

The canonical current can be directly determined from Eq.~\eqref{eqn:big_omega}:
\begin{equation} \label{eqn:big_canonical}
  \bs{\mc E}_\xi \{\delta_1 \bs \Phi, \delta_2 \bs \Phi\} = \bs w \{\delta_1 \bs g, \lie_\xi \delta_2 \bs g\} \cdot \bs \epsilon + \alpha^2 \tilde{\bs w} \{\delta_1 {\bs f}, \lie_\xi \delta_2 {\bs f}\} \cdot \tilde{\bs \epsilon}.
\end{equation}
The effective source current, on the other hand, can be determined from Eqs.~\eqref{eqn:big_C} and~\eqref{eqn:big_E}:
\begin{equation}
  \begin{split}
    (\mc S_\xi)_{abc} \{\delta_1 \bs \Phi, \delta_2 \bs \Phi\} = 2 \xi^d \Big[&(\ord{2} G^{ef} \{\delta_1 \bs g, \delta_2 \bs g\} + m^2 \ord{2}{\mc V}^{ef} \{\delta_1 \bs \Phi, \delta_2 \bs \Phi\}) g_{fd} \epsilon_{eabc} \\
    &+ (\alpha^2 \ord{2}{\tilde G}^{ef} \{\delta_1 \bs f, \delta_2 \bs f\} + m^2 \ord{2}{\tilde{\mc V}}^{ef} \{\delta_1 \bs \Phi, \delta_2 \bs \Phi\}) f_{fd} \tilde \epsilon_{eabc}\Big].
  \end{split}
\end{equation}
Moreover, this equation only holds assuming that the zeroth- and first-order equations of motion hold (so that no variations need to be taken of the volume forms $\bs \epsilon$ and $\tilde{\bs \epsilon}$).
To compute the effective source current, one needs to compute $\ord{2}{\mc V}^{ef}$ and $\ord{2}{\tilde{\mc V}}^{ef}$, both of which are quite lengthy.
Since the canonical current is equivalent, when the equations of motion hold and up to a boundary term, we propose that the canonical current is the ``best'' conserved current to use for problems in bigravity.

We next note the following property of this theory: suppose that $\xi^a$ is a Killing vector of the metric $g_{ab}$, so that $\lie_\xi g_{ab} = 0$.
One can show that the equations of motion then imply that (in general) $\lie_\xi f_{ab} = 0$ as well, as follows: by Eq.~\eqref{eqn:lie_var}, we have that applying a Lie derivative with respect to $\xi^a$ is a type of variation, and so Eqs.~\eqref{eq:FieldEquations_Linearized} and~\eqref{eqn:var_V} yield
\begin{equation}
    \ord{1} G^{ab} \{\lie_\xi \bs g\} + m^2 \left(-\mc V^{ac} g^{bd} \lie_\xi g_{cd} + g^{bc} \ord{1}{\mc U}^a{}_{cd}{}^e \lie_\xi S^d{}_e\right) = 0.
\end{equation}
However, if $\lie_\xi g_{ab} = 0$, this equation then implies that $\lie_\xi S^a{}_b = 0$ as well (so long as $\ord{1}{\mc U}^a{}_{bc}{}^d$, considered as a matrix, has full rank\footnote{We suspect that, for generic $\beta_n$ and $S^a{}_b$, both $\ord{1}{\mc U}^a{}_{bc}{}^d$ and $\ord{1}{\tilde{\mc U}}^a{}_{bc}{}^d$ have full rank.
    If this is not the case, it would imply, from Eqs.~\eqref{eq:FieldEquations_Linearized} and~\eqref{eqn:var_V}, that the equations of motion for $\delta g_{ab}$ and $\delta f_{ab}$ will be decoupled for certain values for $\beta_n$ and $S^a{}_b$.
    A family of solutions where the Killing vectors are not the same, and therefore this matrix cannot have full rank, is given in~\cite{Torsello:2017ouh}.}).
The only way that this can occur is if $\lie_\xi f_{ab} = 0$; a similar argument shows that $\lie_\xi f_{ab} = 0$ implies that $\lie_\xi g_{ab} = 0$.
As such, we do not need to distinguish between $\xi^a$ being a Killing vector for one metric or the other.

We now specialize to proportional backgrounds, as described above in Sec.~\ref{subsec:BiGravLinearizedMetrics}.
In this case, we find that
\begin{equation} \label{eqn:prop_omega}
    \bs \omega\{\delta_1 \bs \Phi, \delta_2 \bs \Phi\} = \hat{\bs w} \{\delta_1 \bs \Phi, \delta_2 \bs \Phi\} \cdot \bs \epsilon,
\end{equation}
where
\begin{equation} \label{eqn:w_hat}
    \hat w^a \{\delta_1 \bs \Phi, \delta_2 \bs \Phi\} \equiv P^{abcdef} \left[\delta_1 g_{bc} \nabla_d \delta_2 g_{ef} + (\alpha/c)^2 \delta_1 f_{bc} \nabla_d \delta_2 f_{ef} - (\interchange{1}{2})\right].
\end{equation}
Moreover, the symplectic current is still diagonal in the basis of $\delta \gamma_{ab}$ and $\delta \phi_{ab}$, much like it was for $\delta g_{ab}$ and $\delta f_{ab}$: using the fact that
\begin{equation} \label{eqn:undiagonalize}
    \delta g_{ab} = \frac{\delta \gamma_{ab} - 2c \alpha^2 \delta \phi_{ab}}{1 + c^2 \alpha^2}, \qquad \delta f_{ab} = \frac{c^2 \delta \gamma_{ab} + 2c \delta \phi_{ab}}{1 + c^2 \alpha^2},
\end{equation}
we find that
\begin{equation} \label{eqn:w_hat_eigenstates}
    \hat w^a \{\delta_1 \bs \Phi, \delta_2 \bs \Phi\} = \frac{P^{abcdef}}{1 + c^2 \alpha^2} \left[\delta_1 \gamma_{bc} \nabla_d \delta_2 \gamma_{ef} + 4 \alpha^2 \delta_1 \phi_{bc} \nabla_d \delta_2 \phi_{ef} - (\interchange{1}{2})\right].
\end{equation}
Note, moreover, that the structure is identical to what appeared in Eq.~\eqref{eqn:w_hat} for each copy.

One property of the canonical current, coming from Eqs.~\eqref{eqn:w_hat_eigenstates}, is that it is completely independent of the mass $m_{\rm FP}$.
This is somewhat unexpected, since, for a massive scalar field, the usual definition of the conserved current associated with a Killing vector, $T_{ab} \xi^b$, will be dependent on the mass.
However, as we show in Appendix~\ref{app:scalar}, it is \emph{also} the case that the canonical current for a massive scalar field is independent of the mass.
For the massive scalar field, the mass shows up in $T_{ab} \xi^b$ due to the equations of motion, which must hold in order for it to be proportional to the canonical current (up to a boundary term).
Similarly, we show that, in the geometric optics limit, the dependence of the stress-energy tensor on the mass goes away, again suggesting that this mass-dependence is not a crucial feature of this conserved current.
Note, however, that in applications of the canonical current, one typically needs to use the equations of motion, which \emph{are} mass-dependent---similarly, in geometric optics, while the conserved current is independent of the mass, the geodesics that are followed by rays are no longer null.

Using Eqs.~\eqref{eqn:canonical}, \eqref{eqn:prop_omega}, and~\eqref{eqn:w_hat_eigenstates}, one can compute the canonical current in terms of the fields $\delta \gamma_{ab}$ and $\delta \phi_{ab}$ (setting, for simplicity, $\delta_1 = \delta_2 = \delta$).
Assuming that one has computed a quadrupole-formula-like expression for these fields [using Eqs.~\eqref{eqn:eigenstate_eom1} and~\eqref{eqn:eigenstate_eom2}], this would allow one to compute the flux of the canonical current through a surface surrounding some compact source, such as that which appears in Fig.~\ref{fig:Radiation}.
In the case where $\xi^a$ is a timelike vector field at infinity, this provides a notion of the ``energy that the source is radiating.''
Note that, \emph{a priori}, there is no reason why this quantity is necessarily related to changes in, say, the binding energy of a compact body---this type of ``flux-balance law'' needs to be \emph{proven}.
However, under the assumption that this flux-balance law holds, the flux of canonical energy can be used to estimate the evolution of a binary merger in bigravity.

Such a flux-balance law is certainly plausible: for example, the canonical current occurs in the flux-balance law for conserved quantities in the first-order scalar self-force~\cite{Grant2022}, and the effective source current (and a straightforward generalization to second-order) occurs in the flux-balance law for second-order gravitational self-force~\cite{second_order20XX}.
Moreover, in Einstein-Maxwell theory, the flux of the canonical current through the event horizon appears in the expression for changes in the mass and spin of a black hole~\cite{Sorce2017}, confirming the notion that it represents a flux of energy/angular momentum in some sense.

\subsection{Asymptotic charges}
\label{sec:asyCharge}
We now (briefly) discuss another approach to conserved quantities which we have not explored in this paper.
In this paper, we have primarily been concerned with conserved currents; integrating such a current over a spacelike surface gives the amount of some conserved quantity that is ``stored in the field'' in some region, and the integral over some timelike or null surface gives the flux of the same conserved quantity.
Using Fig.~\ref{fig:Radiation}, the former is given by the integral over $\Sigma_1$ or $\Sigma_2$, and the latter by the integral over $B_{12}$.
\figRadiation

Another approach is to consider integrals over the boundaries $\partial \Sigma_1$ or $\partial \Sigma_2$.
Such integrals are called ``asymptotic charges'': here ``asymptotic'' refers to the fact that these quantities are computed far from the source, and ``charge'' reflects how they can be computed, much like the electric charge, as an integral over a boundary (in the case of electric charge, the integrand is the electric field).
The simplest example of asymptotic charges considered in vacuum general relativity are the Komar formulae for mass and angular momentum, which are (up to factors) the integrals of the Noether charge in Eq.~\eqref{eqn:Q_def}.
These quantities are related to the Arnowitt-Deser-Misner (ADM) mass and angular momentum; see the discussion in~\cite{Iyer1994} for more details, and an approach which can extend this discussion to more general field theories.
In particular, this shows that the ADM conserved quantities can be understood as conserved quantities associated with asymptotic symmetries, defined in terms of integrals over the sphere at spacelike infinity.
Moreover, they obey an analogue of Hamilton's equations~\cite{Lee1990}: denoting these conserved quantities by $H_\xi$, they satisfy
\begin{equation} \label{eqn:hamiltonian}
    \delta H_\xi = \int_\Sigma \bs \omega(\delta \bs g, \lie_\xi \bs g),
\end{equation}
where $\Sigma$ is a Cauchy surface that extends to infinity.
However, the ADM mass does not evolve---as such, it cannot give useful information about the dynamics of a spacetime.
Instead, one typically considers the Bondi mass, which is defined as an integral over a cross-section of null infinity, and whose evolution depends on an integral over null infinity.

Generalizing the work in~\cite{Iyer1994}, the Wald-Zoupas prescription~\cite{Wald:1999wa} shows that there is a way of understanding the Bondi mass, and other ``conserved quantities'' defined at null infinity, as charges associated with asymptotic symmetries (for a recent pedagogical overview which also discusses relations to other asymptotic charges, see~\cite{Grant2021}).
These quantities now evolve, and their evolution is determined by flux integrals over null infinity.
However, these quantities $\mc Q_\xi (S)$ (depending on a spherical cross-section $S$ of null infinity $\scri$) are no longer ``Hamiltonians'' in the sense of Eq.~\eqref{eqn:hamiltonian}; assuming appropriate falloff conditions as one approaches future timelike and spacelike infinity, one finds instead that, for vacuum general relativity~\cite{Grant2021} (with a similar result for Einstein-Maxwell theory~\cite{Bonga2019})
\begin{equation} \label{eqn:scri_hamiltonian}
    \delta [\mc Q_\xi (S_\infty) - \mc Q_\xi (S_{-\infty})] \equiv \delta \mc F_\xi (\scri) = \int_\scri \bs \omega(\delta \bs g, \lie_\xi \bs g),
\end{equation}
where $S_{\pm \infty}$ are the spheres at $u = \pm \infty$, and $\mc F_\xi (\mc B)$ denotes the ``flux'' of the Wald-Zoupas conserved quantities through a portion of null infinity.
That is, the flux through \emph{all} of null infinity is Hamiltonian.
Moreover, it follows from this equation that, in the case where $\xi^a$ is an exact symmetry,
\begin{equation} \label{eqn:wz_flux_canonical}
    \delta^2 \mc F_\xi (\scri) = \int_\scri \bs{\mc E}_\xi (\delta \bs g, \delta \bs g).
\end{equation}
This is the relationship between the Wald-Zoupas flux and the flux of the canonical current.
Very important for this paper, moreover, is the fact that the methods in~\cite{Wald:1999wa} can be applied, in principle, for any theory that possesses a Lagrangian formulation: it has recently been applied to Einstein-Maxwell theory~\cite{Bonga2019}, Brans-Dicke theory~\cite{Hou2020,Tahura:2020vsa}, and Chern-Simons gravity~\cite{Hou2021}.

In this paper, however, we do not consider applying the Wald-Zoupas procedure to bigravity, and leave a detailed discussion to future work.
The primary reason for this is that (unless the coefficients $\beta_n$ are chosen in a very particular way) bigravity naturally produces solutions which are not asymptotically flat.
In the case of the proportional background, this can be seen from the presence of a cosmological constant in Eq.~\eqref{eqn:prop_einstein}.
More generally, attempting to use the techniques of conformal completion (for an excellent review, see~\cite{Geroch1977}) shows that bigravity does not generally admit asymptotically flat solutions, as the Einstein tensor for either $g_{ab}$ or $f_{ab}$ does not fall off sufficiently rapidly.
Explicitly, if one assumes (as in general relativity) that $g_{ab} = \Omega^{-2} \hat g_{ab}$ and $f_{ab} = \Omega^{-2} \hat f_{ab}$, where $\hat g_{ab}$ and $\hat f_{ab}$ are both smooth at $\Omega = 0$ (which represents the boundary of spacetime), then both $\mc V^{ab}$ and $\tilde{\mc V}^{ab}$ are $O(\Omega^2)$, whereas asymptotic flatness requires that $G^{ab} = O(\Omega^6)$.
While the Wald-Zoupas procedure, in principle, can be applied to any boundary (even in the original paper it was proposed that it could be applied to asymptotically anti-de Sitter spacetimes, and there has been explicit work on finite null surfaces~\cite{Chandrasekaran2018}), there are issues of interpretation in asymptotically de Sitter spacetimes, as their null infinity is a spacelike surface: there is no notion of ``evolution'', and so an integral over null infinity cannot really be called a ``flux''.
While there has been some work in adapting the Wald-Zoupas procedure, and asymptotic charges more generally, to asymptotically de Sitter spacetimes in general relativity~\cite{Anninos2010, Kolanowski2021, Ashtekar2014}, we leave a full discussion in the case of bigravity to future work.

\section{Conclusions and discussion}
This paper aims to study the energy flux of linearized gravitational perturbations in ghost-free bimetric theory (or bigravity). The theory contains two dynamical metrics and has its motivation in the accelerated expansion of the universe~\cite{Schmidt-May:2015vnx} and a self-consistent approach of introducing mass to gravitons.  A special case of bigravity that we consider is where the background metrics are proportional to one another, and the equations of motion decouple to the ones of massive and massless degrees of freedom. We discuss the applicability of several well-known approaches to computing energy flux in bigravity: the Isaacson method~\cite{Isaacson:1968hbi,Isaacson:1968zza}, Wald-Zoupas prescription~\cite{Wald:1999wa}, and a canonical current~\cite{Lee1990, Burnett1990}. We find that the canonical current approach is the simplest for the problem of bigravity.

A gauge-invariant canonical current (equivalent to an effective source current up to a boundary term when the equations of motion hold) can be constructed directly from the Lagrangian in any diffeomorphism covariant theory through symplectic currents. We explicitly express the canonical current in terms of variations of the background metrics for generic bigravity solutions and also for the case where the background metrics are proportional. Interestingly, the mass of the massive degrees of freedom does not appear in the canonical current for the proportional case. This is consistent with the case of a massive scalar field in GR, for which the mass also does not enter the canonical current. By choosing the vector field to be a timelike Killing vector at infinity, one can compute the energy flux by integrating the canonical current over a timelike (or asymptotically null) surface.

Next, we summarize the issues with the Isaacson stress-energy tensor and Wald-Zoupas prescription. The standard method of deriving GW stress-energy tensor is the Isaacson approach which poses several problems when studied in bigravity. In standard GR, a short wavelength averaging is necessary to make the Isaacson stress-energy gauge invariant. Bigravity has additional degrees of freedom and inherently contains multiple length scales such as the characteristic curvature length of the auxiliary metric, Compton wavelength of massive gravitons, and the size of the universe, as it is a theory motivated by cosmology. Because of such various length scales involved, the requirements for short-wavelength averaging are ambiguous (see Fig.~\ref{fig:Wavelength}). Furthermore, the stress-energy tensor is not conserved in bigravity without the averaging. On the other hand, the Wald-Zoupas prescription of ``conserved charges'' associated with asymptotic symmetries requires conformal completion of the spacetime and is usually considered for asymptotically flat spacetimes~\cite{Wald:1999wa} (although there are cases which are not asymptotically flat where it or similar procedures have been applied~\cite{Chandrasekaran2018, Anninos2010}). Bigravity admits asymptotically flat solutions only for a very specific choice of parameters in the proportional background, and for general solutions, the usual conformal completion does not give a solution that is asymptotically flat. As such, one should fully investigate the asymptotics of this theory; insights from the asymptotically de Sitter case in general relativity studied, for example, in~\cite{Ashtekar2014, Ashtekar2015a, Ashtekar2015b, Ashtekar2019} may prove useful.

Finally, let us mention possible future prospects of our work. 
As already mentioned above, the energy content of GWs in bigravity could be studied using the Wald-Zoupas formalism, though this would also require a more rigorous analysis of the asymptotic structures admitted by bigravity.  This can be specialized to the tuned sector admitting asymptotically-flat solutions, but ideally we would generalize to the cosmological setting.  The main application of this work will be to compute GW energy flux radiated by a compact binary source, and subsequently the gravitational waveforms as the binary shrinks due to radiation reaction.

\subsection*{Sketch of flux balance calculation}

Here we can provide a sketch of how this calculation may proceed,
which somewhat parallels the post-Newtonian treatment of compact
binaries in GR.
First, suppose somebody has successfully handled the near zone
generation problem, and can give us linearized radiative fields
$\delta \gamma_{ab}$ and $\delta \phi_{ab}$ in the far
zone, on some stationary background, which determine $\delta \bs \Phi_A = (\delta g_{ab}, \delta f_{ab})$ by Eq.~\eqref{eqn:undiagonalize}.  The energy flux through some
cylinder $B_{12}$ (as in Fig.~\ref{fig:Radiation}) is associated to
the Killing vector field generating the stationarity, $\xi^a = (\partial/\partial t)^a$ as
normalized by some family of observers.  This energy flux would be
computed by integrating the canonical current 3-form,
\begin{align}
  \label{eq:int-dE/dt/dOmega}
  \int_{B_{12}} \frac{dE_{\text{GW}}}{dt d\Omega} dt d\Omega =
  -\int_{B_{12}} \bs{\mc E}_{\xi}(\delta \bs \Phi, \delta \bs \Phi)
  \,.
\end{align}
If we further assume a bidiagonal background, we have that
\begin{align}
  \int_{B_{12}} \bs{\mc E}_{\xi}(\delta \bs \Phi, \delta \bs \Phi) =
  \int_{B_{12}} \hat{n}_{a} \frac{P^{abcdef}}{1 + c^2 \alpha^2}
  \left[
    \delta \gamma_{bc} \nabla_d \delta \dot{\gamma}_{ef} - \delta \dot{\gamma}_{bc} \nabla_d \delta \gamma_{ef}
    +
    4 \alpha^2 (\delta \phi_{bc} \nabla_d \delta \dot{\phi}_{ef} - \delta \dot{\phi}_{bc} \nabla_d \delta \phi_{ef})
  \right]
  dt d\Omega
  \,,
\end{align}
where $\hat{n}^{a}$ is the outward-pointing unit normal to $B_{12}$ (as measured by
$g_{ab}$), and overdots denote time derivatives,
i.e. $\delta\dot{\gamma}_{ab} \equiv \lie_{\xi} \delta\gamma_{ab}$.
In the case where $\delta \gamma_{ab}$ is in the transverse-traceless gauge, one can show that the sum of the terms quadratic in $\delta\gamma_{ab}$ are negative-definite; the same is true for the gauge-invariant $\delta\phi_{ab}$, so the entire right hand side is negative-definite. This reflects the fact that mass is lost by the radiation of gravitational waves.

Note that the $\delta\gamma_{ab}$ terms alone and the
$\delta\phi_{ab}$ terms alone each look like the canonical energy flux as
computed in GR, except (i) they have been scaled differently, and (ii)
$\delta\phi_{ab}$ is a massive field.  Notice that in the eikonal limit, we
will have $\cd_{a} \to ik_{a}$, but that crucially the wavevectors
$k_{a}, k'_{a}$ for the massless and massive field are different.

In the context of a binary inspiral, a first calculation could take
the adiabatic approximation, and thus take $\delta\gamma_{ab}$ and
$\delta\phi_{ab}$ to be periodic with the ``instantaneous'' orbital period
of a quasicircular binary.

We now return to the near zone, the more challenging aspect of such a
calculation.  From separation of scales, we expect to be able to treat
a bound compact binary system (and the GWs it emits) within the
framework of EFT~\cite{Porto:2016pyg}.
First, we want to ``skeletonize'' each compact object, so that we can
replace it with an effective point particle coupled to one or both of
the metrics (and derivatives thereof).  This starts with a
strong-field calculation of, e.g., a neutron star or black hole in this
theory, and then we match in the far zone to the solution from an effective action for a point particle $A$ on world line $\gamma_{A}$ with coordinates
$z_{A}^{a}(\tau_{A})$,
\begin{align}
  S_{\text{pp,eff}}[z_{A}, \bs g, \bs f] = \sum_{A} \int_{\gamma_{A}} -m_{A}(\bs g, \bs f, \ldots) \,d\tau_{A}
  =
  \sum_{A} \int -m_{A}(\bs g, \bs f, \ldots) \delta_{4}(x, z_A) \sqrt{-\det g} \, d^{4}x d\tau_{A}
  \,.
\end{align}
Here $\delta_{4}(x,z)$ is an invariant four-dimensional Dirac
distribution~\cite{Poisson:2011nh}, and $m_{A}(\ldots)$ is a
collection of world-line operators (Wilson coefficients) arising from
matching.  In GR, and assuming the so-called effacement
principle~\cite{Will:2014kxa}, this operator is simply a constant
(finite size effects introduce couplings like
$m_{A} \supset C_{E} E_{ab}(x)E^{ab}(x)$ with $E_{ab}$ being the
electric part of the Weyl tensor~\cite{Porto:2016pyg}).

Next, we create a binary with our two effective point particles
interacting with the metric(s), and thus indirectly with each other.
Split each metric into two fields by the wavelengths of modes,
\begin{align}
  g_{ab} = g_{ab}^{\text{pot}} + g_{ab}^{\text{rad}}
  \,,
\end{align}
and similarly for $f_{ab}$, where the ``potential'' modes
$g_{ab}^{\text{pot}}$ are long wavelength and non-radiative.  Solve for the
potential modes.  Presumably, the massive degree of freedom will have
a Yukawa-like solution, which will add to the usual leading-order
Keplerian potential.
Integrate out the potential modes from the action, to arrive at an
effective action of two point particles interacting directly with each
other, and still coupled to radiative modes,
\begin{align}
  \exp(i S_{\text{eff}}[z_{1}, z_{2}, \bs g^{\text{rad}}, \bs f^{\text{rad}}])
  =
  \int \mathcal{D} \bs g^{\text{pot}} \mathcal{D} \bs f^{\text{pot}}
  \exp(i S_{\text{field}}[\bs g, \bs f] + i S_{\text{pp,eff}}[z_{1}, z_{2}, \bs g, \bs f])
  \,.
\end{align}
Here $S_{\text{field}}$ is the action for bigravity.  $S_{\text{eff}}$
will include an effective interaction potential between the two bodies
(in GR, this procedure yields the Einstein-Infeld-Hoffmann
Lagrangian).  Now we can extract the binding energy $E_\text{bind}(r)$
and orbital frequency relation $\omega(r)$ on a circular orbit.  A
reasonable guess is that $E_\text{bind}(r)$ will include a Yukawa term
after integrating out the massive potential mode. However since the
Compton wavelength of the massive graviton is extremely long compared
to the size of the system, this may well be approximated by a
Kepler potential with a different effective Newton's constant.

The coupling to radiative modes remaining in $S_{\text{eff}}$ allows
us to compute the radiated GWs in terms of the orbit.  Now assuming a
flux-balance law, these ingredients are enough to compute the slow
inspiral for a quasicircular binary, by asserting that
\begin{align}
  \dot{E}_\text{GW} =
  \int \frac{dE_{\text{GW}}}{dt d\Omega} d\Omega
  &=-\frac{d}{dt}E_\text{bind}(r) =
  -\frac{dE_\text{bind}}{dr}\frac{dr}{dt}
  \,,
\end{align}
where $dE_{\text{GW}}/dt/d\Omega$ appeared in Eq.~\eqref{eq:int-dE/dt/dOmega},
coming from the canonical current.
However, we leave implementing this sketch to future work.

\acknowledgments

We thank
B. Bonga,
\'E. Flanagan,
and
K. Prabhu
for
illuminating discussions,
and E. Poisson, H. Yang, A. Tolley, and D.~Nichols for valuable feedback on an earlier version
of this manuscript.
The work of L.C.S. was partially supported by NSF CAREER Award PHY-2047382.
S.T. was supported by Ed Owens family foundation, Natural Sciences and Engineering Research Council of Canada and partially by Perimeter Institute for Theoretical Physics. Research at Perimeter Institute is supported in part by the Government of Canada through the Department of Innovation, Science and Economic Development Canada and by the Province of Ontario through the Ministry of Colleges and Universities.
A.S. is funded by the Center for Gravitational Waves and Cosmology at West Virginia University.

\appendix

\section{Scalar field example} \label{app:scalar}

In this appendix, we consider the example of a massive scalar field, and derive two properties of this theory.
First, we show that, while the canonical current only contains information from the ``kinetic'' part of the Lagrangian, the stress-energy current is still related to it by a boundary term, although the stress-energy current also contains information about the ``potential'' part of the Lagrangian.
This feature is shared by bigravity, but it is far less clear in that context how this arises.
Second, we consider the geometric optics limit of this theory, and show that the mass term does not contribute to the leading-order average of the stress-energy tensor.

To prove the first result, we start with the Lagrangian for this theory:
\begin{equation}
  \bs L = -\frac{1}{2} \bs \epsilon \left[g^{ab} (\nabla_a \phi) (\nabla_b \phi) + m^2 \phi^2\right].
\end{equation}
Once again, note that $m$ has units of inverse length.
Varying the Lagrangian gives
\begin{equation}
  \begin{split}
    \delta \bs L &= -\left[g^{ab} (\nabla_b \phi) (\nabla_a \delta \phi) + m^2 \phi \delta \phi\right] \bs \epsilon \\
    &= \delta \phi \left(g^{ab} \nabla_a \nabla_b \phi - m^2 \phi\right) \bs \epsilon - \nabla_a (\delta \phi g^{ab} \nabla_b \phi) \bs \epsilon \\
    &\equiv \bs E \delta \phi + \ud \bs \theta \{\delta \phi\},
  \end{split}
\end{equation}
where
\begin{equation}
  \bs E = \bs \epsilon \left(\Box \phi - m^2 \phi\right)
\end{equation}
and
\begin{equation}
  \bs \theta \{\delta \phi\} = \bs v \{\delta \phi\} \cdot \bs \epsilon, \qquad v^a \{\delta \phi\} \equiv -g^{ab} \delta \phi \nabla_b \phi,
\end{equation}
and so
\begin{equation} \label{eqn:scalar_omega}
  \bs \omega \{\delta_1 \phi, \delta_2 \phi\} = \bs w \{\delta_1 \phi, \delta_2 \phi\} \cdot \bs \epsilon, \qquad w^a \{\delta_1 \phi, \delta_2 \phi\} \equiv g^{ab} \left[\delta_1 \phi \nabla_b \delta_2 \phi - (\interchange{1}{2})\right].
\end{equation}
We therefore have that the canonical current is
\begin{equation} \label{eqn:scalar_E}
  \bs{\mc E}_\xi \{\delta_1 \phi, \delta_2 \phi\} = \bs w \{\delta_1 \phi, \lie_\xi \delta_2 \phi\} \cdot \bs \epsilon.
\end{equation}

We now consider the stress-energy tensor, which is obtained (up to unimportant numerical factors) by taking a variation of the Lagrangian with respect to $g_{ab}$:
\begin{equation} \label{eqn:scalar_T}
  T^{ab} \bs \epsilon \equiv \frac{\delta \bs L}{\delta g_{ab}} = \frac{1}{2} \left\{g^{ac} g^{bd} (\nabla_c \phi) (\nabla_d \phi) - \frac{1}{2} g^{ab} \left[g^{cd} (\nabla_c \phi) (\nabla_d \phi) + m^2 \phi^2\right]\right\} \bs \epsilon,
\end{equation}
and so
\begin{equation}
  (j_\xi)^a \equiv T^{ab} g_{bc} \xi^c = \frac{1}{2} \left\{g^{ab} \xi^c (\nabla_b \phi) (\nabla_c \phi) - \frac{1}{2} \xi^a \left[g^{bc} (\nabla_b \phi) (\nabla_c \phi) + m^2 \phi^2\right]\right\}.
\end{equation}
The stress-energy current is then related to $(j_\xi)^a$ by
\begin{equation}
  \bs{\mc J}_\xi \equiv \bs j_\xi \cdot \bs \epsilon.
\end{equation}

Our goal is to show that, up to constant factors, a boundary term, and equations of motion, the stress-energy current is the same as the canonical current.
That is, we need that
\begin{equation}
  \bs{\mc E}_\xi \{\phi, \phi\} = C \bs{\mc J}_\xi + \ud \bs{\mc Q}
\end{equation}
holds when the equations of motion hold.
Note, first, that this implies that
\begin{equation} \label{eqn:scalar_relation}
  w^a \{\phi, \lie_\xi \phi\} = C (j_\xi)^a + \nabla_b Q^{ab},
\end{equation}
where $Q^{(ab)} = 0$ and
\begin{equation}
  \mc Q_{ab} = \frac{1}{2} Q^{cd} \epsilon_{cdab}.
\end{equation}
Since it is easier, we will work entirely with the vector form of these equations.

We start with inspecting terms in $w^a \{\phi, \lie_\xi \phi\}$, which is given by
\begin{equation}
  w^a \{\phi, \lie_\xi \phi\} = g^{ab} \left[-\xi^c (\nabla_b \phi) (\nabla_c \phi) + (\nabla_b \xi^c) \phi \nabla_c \phi + \xi^c \phi \nabla_b \nabla_c \phi\right]
\end{equation}
The first term involves $(\nabla_a \phi) (\nabla_b \phi)$, which also appears in the stress-energy current.
The second and third terms, however contain $\nabla_a \xi^b$ and $\nabla_a \nabla_b \phi$, and these types of terms do \emph{not} appear in the stress-energy current.
For the second term, we use the fact that $\nabla_{(a} \xi_{b)} = 0$ to show that
\begin{equation}
  g^{ab} (\nabla_b \xi^c) \phi \nabla_c \phi = -g^{bc} (\nabla_c \xi^a) \phi \nabla_b \phi = -\nabla_c (g^{bc} \xi^a \phi \nabla_b \phi) + \xi^a \left[g^{bc} (\nabla_b \phi) (\nabla_c \phi) + \phi \Box \phi\right].
\end{equation}
For the third term, we can use the fact that $\nabla_a \xi^a = 0$ to show that
\begin{equation}
  g^{ab} \xi^c \phi \nabla_b \nabla_c \phi = \nabla_c (g^{ab} \xi^c \phi \nabla_b \phi) - g^{ab} \xi^c (\nabla_b \phi) (\nabla_c \phi).
\end{equation}
As such, we have that
\begin{equation}
  w^a \{\phi, \lie_\xi \phi\} = -2 g^{ab} \xi^c (\nabla_b \phi) (\nabla_c \phi) + \xi^a [g^{bc} (\nabla_b \phi) (\nabla_c \phi) + \phi \Box \phi] - \nabla_b (2 \xi^{[a} g^{b]c} \phi \nabla_c \phi).
\end{equation}
Using the equations of motion, $\Box \phi = m^2 \phi$, we find that Eq.~\eqref{eqn:scalar_relation} holds, with
\begin{equation}
  C = -4, \qquad Q^{ab} = -2 \xi^{[a} g^{b]c} \phi \nabla_c \phi.
\end{equation}

We now prove the second result.
To do so, we start with the geometric optics ansatz for a massive scalar field:
\begin{equation}
    \phi = A \Re[e^{-i\theta}],
\end{equation}
where $A$ is a slowly varying amplitude and $\theta$ a rapidly varying phase.
Applying this approximation to the equations of motion yields, at leading order,
\begin{equation} \label{eqn:eikonal}
    k^a k_a + m^2 = 0,
\end{equation}
where $k_a = \nabla_a \theta$ is the wavevector.
Applying the Brill-Hartle average, we find that, to leading order,
\begin{equation}
    \left\langle T^{ab}\right\rangle = \frac{1}{4} A^2 k^a k^b,
\end{equation}
where we have used Eqs.~\eqref{eqn:scalar_T} and~\eqref{eqn:eikonal}.
The mass term therefore does not contribute to the average of $T^{ab}$.

Note that there is also another way of deriving this result, which should also be applicable to any reasonable notion of geometric optics in bigravity.
We have already shown that $\bs J_\xi$ and $\bs{\mc E}_\xi$ are the same, up to overall constants and a boundary term, for any Killing vector.
The canonical current, applying the geometric optics limit and the Brill-Hartle average, is given by
\begin{equation}
    \left\langle\mc E_\xi\{\phi, \phi\}\right\rangle = -A^2 k^a k^b \xi_b,
\end{equation}
where we have used Eqs.~\eqref{eqn:scalar_omega} and~\eqref{eqn:scalar_E}.
Since the canonical current does not contain any information about the mass of the scalar field, the geometric optics limit clearly does not either.
Now, consider the geometric optics limit of the boundary term $\ud \bs{\mc Q}$.
There are two cases: first, suppose the $\bs{\mc Q}$ is slowly varying.
In that case, $\ud \bs{\mc Q}$ is suppressed in the geometric optics limit.
Instead, suppose then that $\bs{\mc Q}$ is rapidly varying.
In that case, while $\ud \bs{\mc Q}$ would be not suppressed, it would \emph{also} be rapidly varying, and therefore be suppressed by the Brill-Hartle average.
In either case, therefore, $\ud \bs{\mc Q}$ cannot contribute: the Brill-Hartle average is insensitive to boundary terms.
A similar result appears in~\cite{Isaacson:1968zza, Grant2020} for linearized gravity.
This suggests that, should a sensible notion of these ideas exist in bigravity, the Brill-Hartle average of the canonical current would be the same as the current constructed from that theory's corresponding notion of an Isaacson stress-energy tensor, which would be the Brill-Hartle average of the effective source current.

\bibliographystyle{JHEP}
\bibliography{bigravBib}

\end{document}